\documentclass[aps,prl,reprint,twocolumn,raggedbottom]{revtex4-1}
\usepackage{graphicx}
\usepackage{amsmath}
\usepackage{amssymb}
\usepackage{braket}
\usepackage[colorlinks,allcolors=blue]{hyperref}
\usepackage{hypcap}
\usepackage{soul}
\newcommand{\squig}{{\raise.17ex\hbox{$\scriptstyle\sim$}}}
\newcommand{\pref}[2]{\hyperref[#1]{\ref{#1}#2}}

\begin{document}

\title{Polarization spectroscopy of atomic erbium in a hollow cathode lamp}
\author{Jackson Ang'ong'a and Bryce Gadway}
\affiliation{Department of Physics, University of Illinois at Urbana-Champaign, Urbana, IL 61801-3080, USA}
\date{\today}

\begin{abstract}
In this work we perform polarization spectroscopy of erbium atoms in a hollow cathode lamp (HCL) for the stabilization of a diode laser to the 401-nm transition. We review the theory behind Doppler-free polarization spectroscopy, theoretically model the expected erbium polarization spectra, and compare the numerically calculated spectra to our experimental data. We further analyze the dependence of the measured spectra on the HCL current and the peak intensities of our pump and probe lasers to determine conditions for optimal laser stabilization.
\end{abstract}

\maketitle

As research on laser-cooled gases has matured, the variety of atomic and molecular species under investigation has continued to expand, opening up new directions in quantum simulation research~\cite{bloch,lahaye:dipolar_review_2009}, precision measurements of fundamental quantities~\cite{ACME:EDM2014}, and quantum information studies~\cite{HolmiumComputing}. In the context of quantum simulation, there has been much interest over the past decade in the laser cooling of atomic species with large magnetic dipole moments, which can play host to nontrivial long-ranged dipolar interactions~\cite{lahaye:dipolar_review_2009}. Interest and activity in this area have been fueled by tremendous advances in the laser cooling and trapping of chromium~\cite{ChromiumMOT} and erbium~\cite{Ban:05,mclelland2,McClelland-Erb}, as well as by the production of dipolar atomic quantum gases of chromium~\cite{GriesBEC,Cr-Fermi}, dysprosium~\cite{lu:strongly_Dy_2011,lu:quantum_Dy_2012}, and erbium~\cite{aikawa:bose-einstein_Er_2012,Aikawa-Fermi}.

Beginning from the first theoretical study of erbium as a candidate for cooling and trapping~\cite{Ban:05}, experimental studies on frequency stabilization using atomic transition lines in erbium have been performed. The techniques of frequency modulation spectroscopy~\cite{brammer}, where multiple frequency components of a probe beam interfere and produce a demodulated error signal, and modulation transfer spectroscopy~\cite{frisch}, where a background free error signal is generated by an induced four-wave mixing process~\cite{Shirley:82}, have been explored previously. For the near-ultraviolet (near-UV) transitions of the lanthanides, which feature very large natural linewidths, the technique of polarization spectroscopy offers a simplified, modulation-free alternative that can achieve the necessary levels of laser stability. In this work, we explore for the first time the use of polarization spectroscopy of erbium for laser frequency stabilization.

The outline of the paper is as follows: Section~\ref{background} surveys the general properties of atomic erbium and gives a brief introduction to polarization spectroscopy. In Section~\ref{exptsetup} we discuss the experimental setup. We then discuss in Section~\ref{results} the measured spectra, dependence on various experimental parameters, and details of the data analysis. Section \ref{theory} discusses the main theoretical aspects of polarization spectroscopy. We present concluding remarks in Section~\ref{conclusions}.

\section{Background} \label{background}
\subsection{Properties of erbium}

\begin{figure}[t!]
	\includegraphics[width=\columnwidth]{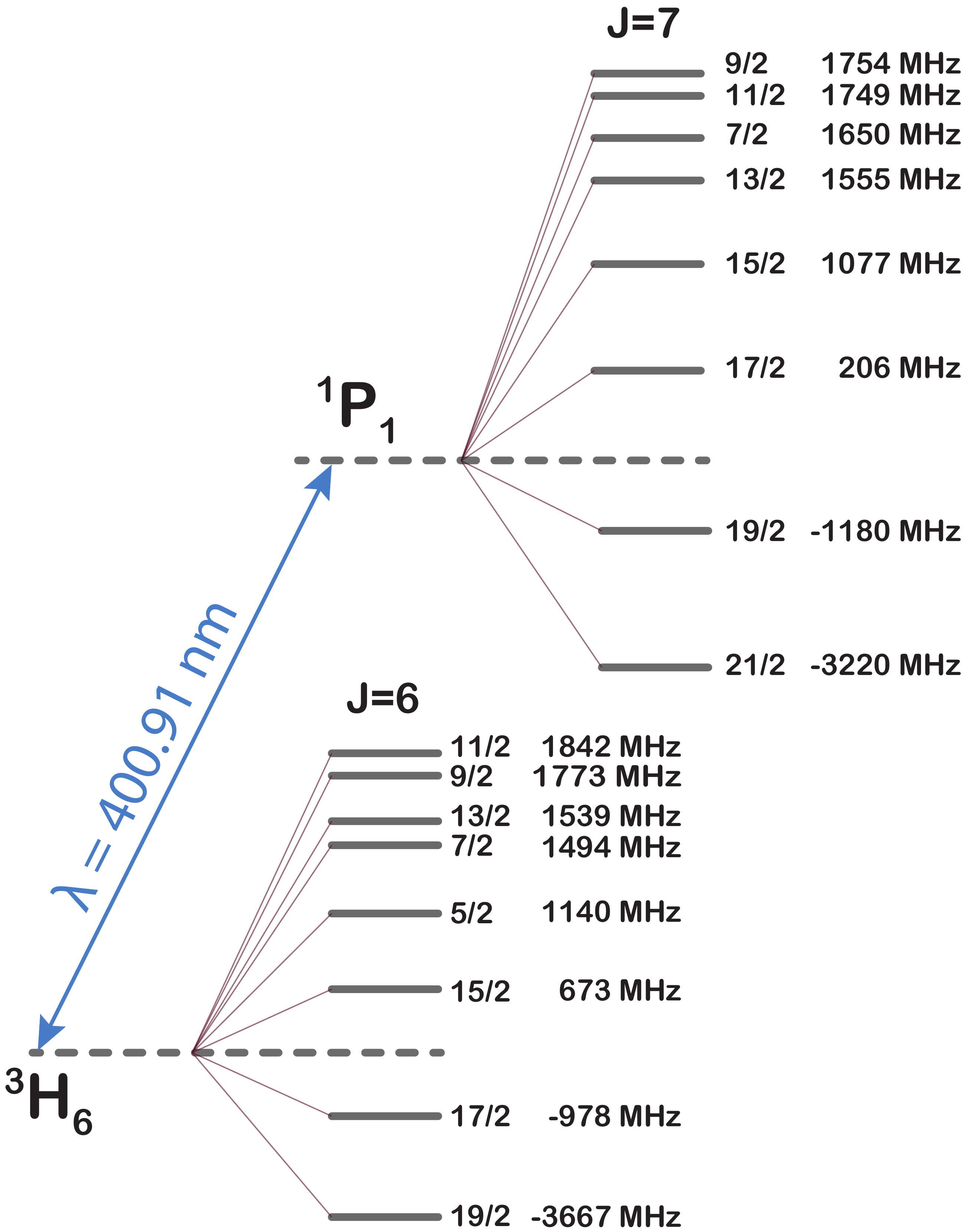}
	\caption{Hyperfine structure of fermionic $^{167}$Er, shown for the $^3H_6$ ground state and the $^1P_1$ excited state (both indicated by dashed lines). The frequency detuning values from the 401-nm transition line are calculated using hyperfine constants  $A_e/h =-100$~MHz and $B_e/h =-3079$~MHz for the $^1P_1$ excited state and  $A_g/h =-120$~MHz and $B_g/h =-4552$~MHz for the $^3H_6$ ground state \cite{frisch}.}
	\label{fig:hfstruct}
\end{figure}

Erbium and other magnetic lanthanide atoms have gained interest in recent years in the study of dipolar quantum gases~\cite{lahaye:dipolar_review_2009}. Due to their submerged-shell electron configuration~\cite{Wybourne}, lanthanides exhibit a large dipole moment ($7 \mu_B$ for erbium, where $\mu_B$ is the Bohr magneton) and support long-ranged dipolar interactions that can dominate over short-ranged collisions. This has made them promising for the study of long-ranged interacting Hubbard models~\cite{Ferlaino-Extended-2015}, dipolar spin systems~\cite{dePaz-Chrom-QM-2013}, and many other novel many-body systems.

Erbium has a melting temperature of $\squig 1800$~K, much greater than that of the alkali metal atom species ($\squig 300-350$~K). Thus, while vapor pressures compatible with laser spectroscopy may be achieved in alkali metal gas cells at or slightly above room temperature, extremely high temperatures ($\gtrsim 1250$~K) would be necessary to achieve similar vapor pressures for the spectroscopy of atomic erbium. This has motivated the use of optogalvanic hollow cathode lamps (HCLs) for laser spectroscopy~\cite{frisch,brammer} on high density, non-thermal samples of erbium atoms produced by sputtering from an erbium cathode. As detailed in Section~\ref{exptsetup}, we employ such an HCL in our polarization spectroscopy experiments.

We explore the near-UV transitions in erbium, occurring at a nominal transition wavelength (in vacuum) of 400.91~nm. The near-UV transitions of erbium and other lanthanide atoms feature large natural linewidths of $\Gamma/2\pi \sim 25-30$~MHz, allowing for the application of strong optical forces. The large linewidths also enable rather straightforward frequency stabilization.

The erbium spectra are expected to show contributions primarily from four bosonic isotopes, with natural abundances of $1.61\%$ for $^{164}$Er, $33.6\%$ for $^{166}$Er, $26.8\%$ for $^{168}$Er and $15.0\%$ for $^{170}$Er, as well as from fermionic $^{167}$Er, with an abundance of $23.0\%$ \cite{frisch}. The bosonic isotopes have nuclear quantum number $I=0$, and thus have no hyperfine structure in either the ground or excited electronic state. This lack of hyperfine structure leads to extremely simple spectral features. On the other hand, the fermionic isotope $^{167}$Er has a nuclear quantum number $I=7/2$ and therefore features a rich hyperfine structure in the $^3H_6$ electronic ground state with angular momentum quantum number $J=6$ and the $^1P_1$ state excited by the 401-nm transition, having $J = 7$ as shown in Fig.~\ref{fig:hfstruct}.

\subsection{Doppler-free polarization rotation spectroscopy}

Laser spectroscopy free from Doppler shifts experienced by the fast-moving erbium atoms in our HCL is enabled through a well-studied
\cite{pearman,harris,Yoshikawa:03,Zhu:14,Do}, modulation-free variant of Doppler-free techniques based on overlapping pump and probe beams.~In the general Doppler-free spectroscopy setup, the lower intensity probe beam features a narrow response in its interaction with the atomic medium at resonance, i.e. when the probe and the higher intensity pump beam interact with the same class of atoms near zero velocity (projected along the axis of beam propagation)~\cite{HoleBurning,SatSpec}.

In Doppler-free polarization spectroscopy~\cite{WiemanPolSpec}, a circularly-polarized pump beam causes an anisotroptic population redistribution of the magnetic sublevels ($|F,m_F\rangle$ states, or $|J,m_J\rangle$ states for the case where $I=0$) in the atomic medium through optical pumping, which induces a rotational birefringence (unequal refractive indices experienced by light with opposite circular polarizations) and circular dichroism (unequal absorption coefficients experienced by light with opposite circular polarizations).~A linearly-polarized probe beam overlapping the pump beam can thus experience a rotation of its polarization due to the different interactions of the right- and left-circular components of the light with the anisotropic atomic medium. This rotation, measured in terms of a difference signal at a balanced photodetector, is then converted to an error signal that can be used for frequency stabilization, as explained in more detail in Section \ref{exptsetup}. 

While such a modulation-free scheme is much simpler than alternative protocols~\cite{frisch}, it does suffer from greater noise and sensitivity to variations of temperature and laser power. However, it should be more than adequate for laser stabilization to the broad near-UV transitions of the lanthanides~\cite{Zhu:14}.
\begin{figure}[t!]
	\includegraphics[width= \columnwidth]{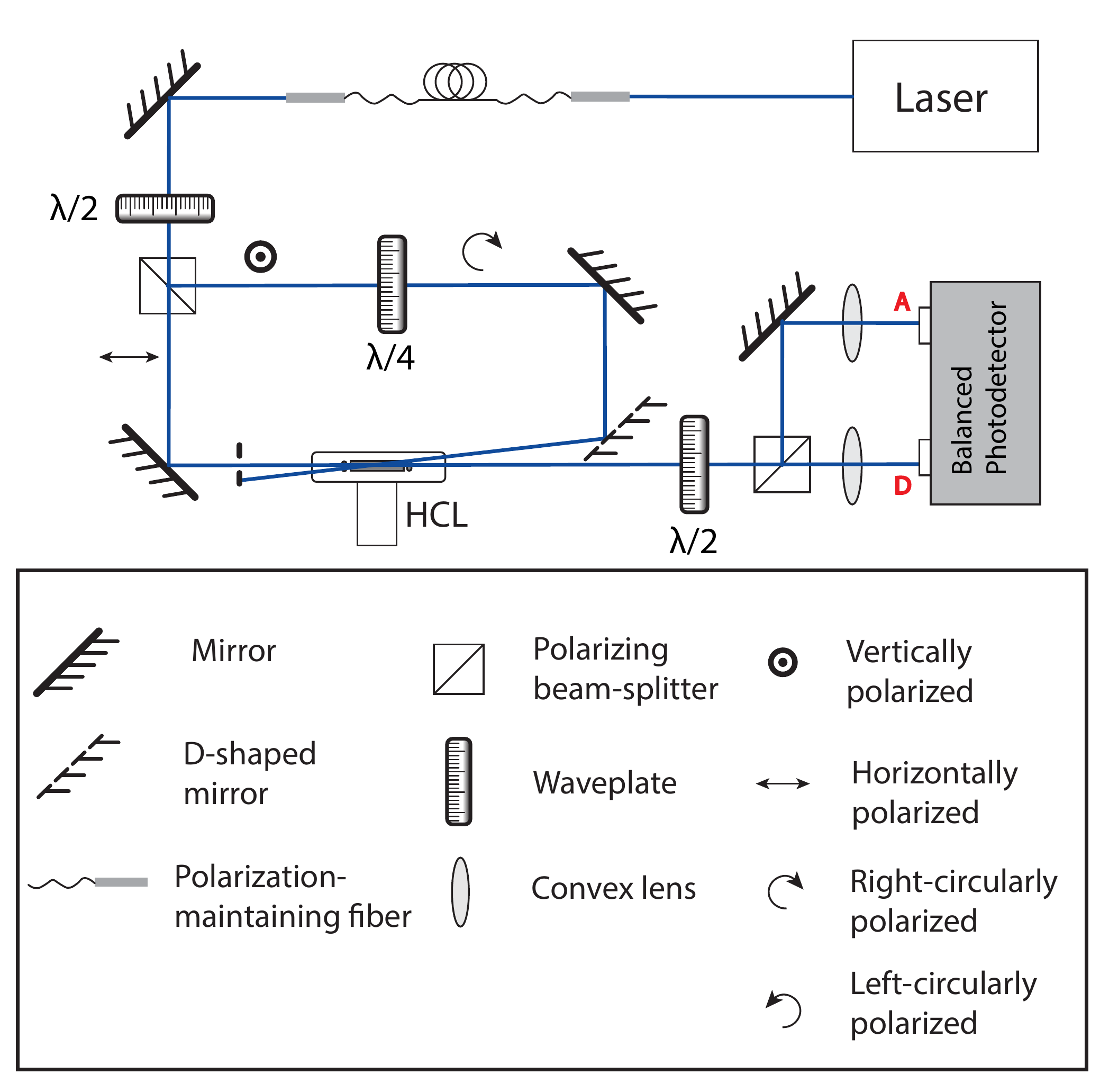}
	\caption{Experimental setup.
		The laser beam is divided into a high power pump beam (reflected at the polarizing beam-splitter (PBS)) and a low power probe beam (transmitted at the PBS). The half-wave plate (HWP) before the first PBS controls the relative powers of the pump and probe beams. The vertically-polarized pump beam encounters a quarter-wave plate (QWP), whose setting controls the polarization of the pump beam (can turn the initial vertical polarization of the pump beam to circular polarization). The D-shaped mirror combines the pump and probe beams in a nearly counter-propagating fashion. The probe beam passes through the HCL and is then split into its anti-diagonal ($A$) and diagonal ($D$) components at the second PBS which are measured at the two ports ($A$ and $D$) of the balanced photodetector. The difference signal between the measurements at these two ports contstitutes the polarization spectroscopy signal.}
	\label{erbsetup}
\end{figure}
\section{Experimental setup}
\label{exptsetup}

The atomic medium used for spectroscopy and laser frequency stabilization is produced by an optogalvanic HCL (Heraeus Model 3QQAYEr) filled with inert argon gas at pressures in the range of 8-15 mbar and containing an erbium cathode. The gas of erbium atoms is produced by applying a large voltage between the anode and the cathode which accelerates ionized filler gas atoms into the erbium cathode, leading to the sputtering of erbium atoms and ions, which decay and can be interrogated by lasers passed through the ``see-through'' hole. For the described measurements, we operate the cathode lamp at HCL current values ranging between 5.5~mA and 10~mA.

We operate a grating-stabilized external cavity diode laser (Toptica Photonics DL pro HP, with LD-0405-0250-1 laser diode) at a nominal wavelength of 401~nm. As shown in Fig.~\ref{erbsetup}, the laser light is fiber-coupled (in a single mode fiber) to the spectroscopy setup, to ensure a Gaussian beam profile. In the spectroscopy setup, the laser beam passes through a half-wave plate (HWP) and a polarizing beam-splitter (PBS). The setting of this first HWP allows us to control the relative powers of the pump (reflected at the PBS) and probe (transmitted through the PBS) beams. The horizontally-polarized probe beam is aligned in a forward-going fashion through the central opening of the HCL. The path of the pump beam features a quarter-wave plate (QWP) that allows for the initial vertical polarization of the reflected light to be either kept linear or transformed into either right- or left-circular polarization (elliptical polarizations for intermediate settings). This pump beam is then overlapped with the probe beam in the central region of the HCL. The use of a D-shaped mirror combines these beams at a small incidence angle, i.e. in a nearly counter-propagating fashion, to enhance the size of the region of overlap.

After propagating through the HCL, the probe beam passes through a HWP and a PBS, and the laser powers in the transmitted and reflected paths are measured at two ports ($A$ and $D$) of a balanced photodetector. The difference signal between the measurements at these two ports constitutes the polarization spectroscopy signal. The setting of the final HWP effectively allows the final PBS to project the probe light onto its diagonal ($D$) and anti-diagonal ($A$) linear polarization components. In the absence of rotation of the incident horizontally-polarized probe induced by erbium atoms in the HCL, the measured powers in the $A$ and $D$ paths will be equal, such that their measured difference signal at the photodetector will be zero. Non-zero difference signals relate to a combination of dichroism and birefringence induced in the erbium atomic medium, due to anisotropic population redistribution by the pump beam. The polarization spectroscopy traces are recorded by capturing the response of this setup as the piezo voltage (and thus laser frequency) is scanned in a symmetric (triangle wave) fashion.

The quoted pump and probe laser powers, relating to the powers in the region of overlap in the HCL, take into account a large loss of \squig19$\%$ at the glass surface of the HCL.
Due to spatial aberrations in the outer portions of the pump and probe beams, we present our measurements in terms of the peak intensity, i.e. $I_{\mathrm{peak}} = 2P_0/\pi \sigma_x \sigma_y$, of the central Gaussian portion of the beam. Here, $P_0$ is the power in the central Gaussian portion (determined from a central fit and the total beam power), $\sigma_x$ is the $1/e^2$ half-width along the horizontal axis (418(1) $\mu$m and 645(2) $\mu$m for the pump and probe beam, respectively, determined by fitting), and $\sigma_y$ is the $1/e^2$ half-width along the horizontal axis (476(1) $\mu$m and 653(2) $\mu$m for the pump and probe beam, respectively).

\begin{figure}[b!]
	\includegraphics[width=\columnwidth]{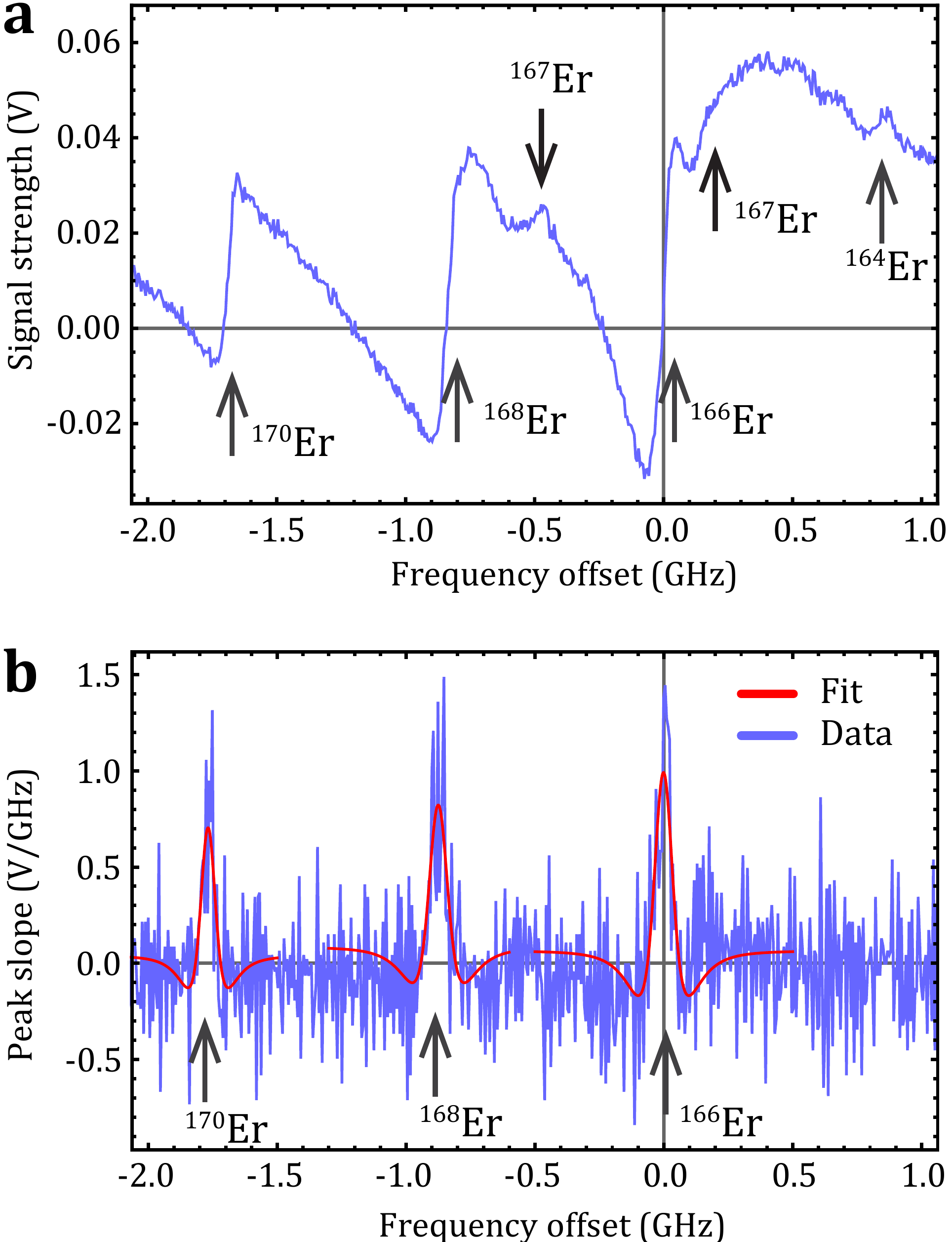}
	\caption{Measured polarization spectroscopy signal and peak slope estimation. (a) Typical polarization spectroscopy signal. The transition associated with $^{166}$Er is taken to be the reference frequency so that  $^{164}$Er, $^{168}$Er and  $^{170}$Er are offset at $849$~MHz, $-840$~MHz and $-1681$~MHz \cite{frisch}, respectively.~Two small features relating to $^{167}$Er can be distinguished as well. (b) Second derivative of Lorentzian fits to extract the signal peak slope associated with the bosonic isotopes. For these measurements, the pump peak intensity is set to 7.9(1)~mW/mm$^2$, the probe peak intensity is 0.535(5)~mW/mm$^2$, and the current applied to the HCL is 5.5~mA.}
	\label{lorentzianfit}
\end{figure}
\section{Results and Discussion}
\label{results}
\begin{figure*}[t!]
	\includegraphics[width=2\columnwidth]{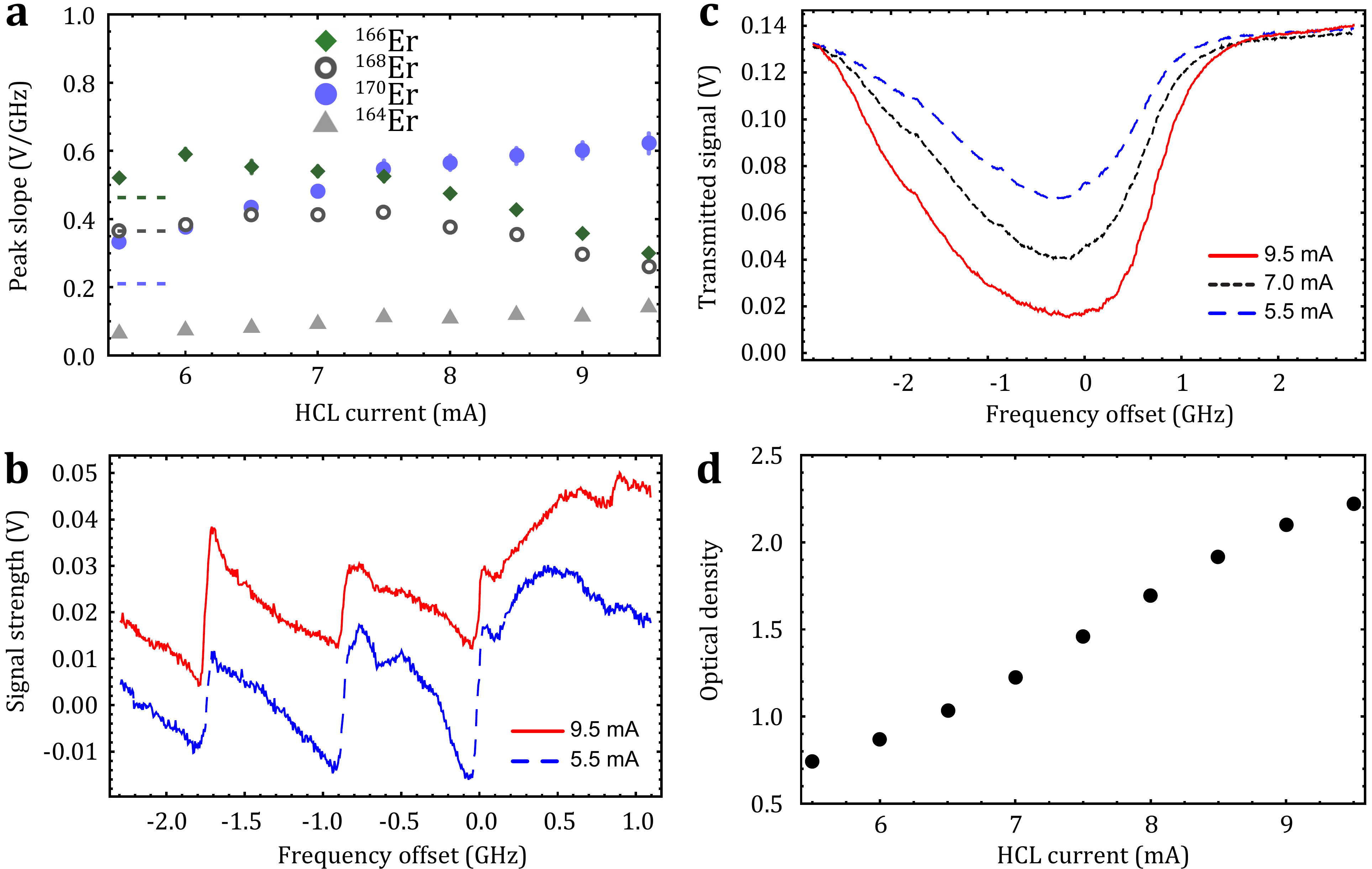}
	\caption{Dependence of polarization spectra on the atomic density, as controlled by the HCL current. (a)~Peak slope as a function of HCL current for the various bosonic isotopes $^{164}$Er, $^{166}$Er, $^{168}$Er, and $^{170}$Er. The horizontal lines at the far left indicate expected peak slope (scaled to match the natural abundances) with the middle gray line matched to the $^{168}$Er peak slope value for an HCL current of 5.5 mA.
		(b)~Polarization spectra for the cases of low (5.5~mA, blue dashed line) and high (9.5~mA, red solid line) applied HCL current. (c)~Transmitted probe signal collected at the A port of the balanced detector. (d)~ Optical density OD~$=-\mathrm{ln} (T_{\mathrm{min}}/T_{\mathrm{max}})$ as a function of HCL current. For all data shown, the pump and probe beam peak intensities were held at 7.9(1)~mW/mm$^2$ and 0.535(5)~mW/mm$^2$, respectively, while the QWP was set so that the pump beam had right-circular polarization. Error bars denote one standard error.
	}
	\label{HCL-plot}
\end{figure*}

One of the main features of relevance to laser stabilization is the slope of the signal as a function of frequency, since a large slope allows for a sizeable response even for small frequency deviations away from resonance.
Figure~\pref{lorentzianfit}{(a)}  shows a typical polarization spectroscopy trace measured from the output of the balanced photodetector. For all measurements taken we set the piezo peak-to-peak scan voltage at 2.1 V. The frequency axis is calibrated using data from Ref.~\cite{frisch}, where the transition associated with $^{166}$Er is taken to be the reference frequency so that  $^{164}$Er, $^{168}$Er and  $^{170}$Er transitions are offset at $849$~MHz, $-840$~MHz and $-1681$~MHz, respectively. Due to broadening and the relatively close spacing of the fermionic and bosonic transitions, the fermionic features are mostly obscured. However, two small features relating to $^{167}$Er can be distinguished, as explained further in Section \ref{theory}.

In this section, we investigate the dependence of this spectroscopy signal slope on parameters such as the HCL current, the pump beam peak intensity, the probe beam peak intensity, and the pump laser polarization. In these later measurements, since the signal takes the form of the derivative of a Lorentzian, we determine the slope of the signal for the three dominant bosonic transitions by fitting the peak height of the discrete derivative curves as in Fig.~\pref{lorentzianfit}{(b)} to the form of the second derivative of a Lorentzian. For each distinct experimental arrangement, six different data curves are recorded and their respective fit-determined maximum polarization signal slope values are calculated. These collections of measured values are used to extract the mean value and standard error of the signal slope for each set of experimental conditions.

\subsection{HCL current}
\label{hclsection}
We first examine how polarization spectra depend on the operating current of the HCL, which controls the number density of erbium atoms in the probing region. The pump and probe beam peak intensities considered in this case are 7.9(1)~mW/mm$^2$ and 0.535(5)~mW/mm$^2$, respectively. At lower HCL current values, the magnitudes of the spectral features associated with the three dominant bosonic isotopes ($^{166}$Er, $^{168}$Er, and $^{170}$Er) are roughly correlated, at least in ordering, with their respective natural abundances. That is, in Fig.~\ref{lorentzianfit} the slope of the feature associated with $^{166}$Er ($33.6\%$ in natural abundance) exceeds that of the less populous $^{168}$Er ($26.8\%$), which in turn exceeds that of the still less populous $^{170}$Er~($15.0\%$). When the HCL current is increased as shown in Fig.~\pref{HCL-plot}{(a)}, the magnitude of the dispersive features are no longer correlated (in ordering) with their natural abundances.

The correspondence between the signal slopes and the isotopic abundances at low HCL currents is emphasized by the dashed lines shown at the left of the plot in Fig.~\pref{HCL-plot}{(a)}, where the middle line is fixed by the slope of the $^{168}$Er transition and the other two are scaled by the relative isotopic abundances. As the HCL current is tuned from 5.5~mA to 6.0~mA, the atomic density in the probing region grows, leading to an increase in the slopes. The slope for the $^{166}$Er transition begins to turn over and then decrease for increasing HCL currents beyond 6.0~mA, the slope of the $^{168}$Er feature turns over and decreases for HCL currents beyond 7.0~mA, and the slope of the $^{170}$Er feature is only increasing over the range of HCL currents explored. The feature relating to the $^{164}$Er transition, mostly indistinguishable for low HCL currents, is observable and has an increasing slope for increasing HCL currents above 7.0~mA.

This observed behavior is due to competing processes that result from growing atomic density. The amount of rotation of the probe beam's linear polarization (induced by the atomic medium) increases for an increasing atomic density, as well as for any increase of the effective length of the interrogated region, which raises the number of scatterers that interact with the probe beam. This increase in the polarization rotation leads  to larger spectral features with larger signal slopes, as is observed for increasing HCL currents when the currents are low. Competing with this is the increase in absorption by the atomic medium which corresponds to a decrease in the probe beam signal transmitted. This leads to a decrease in the size of dispersive signal obtained. Such a trend can explain the behavior observed in Fig.~\pref{HCL-plot}{(a,b)}, where the signal slopes for the more abundant isotopes tend to turn over and decrease at lower HCL currents, with absorption dominating over induced rotation.

Figure \pref{HCL-plot}{(c)} shows the transmitted probe signal, $T(\omega)$, measured at the $A$ port, for HCL currents $5.5$~mA, $7.0$~mA and $9.5$~mA. Since $T(\omega)$ at the $A$ port is approximately equal to $T(\omega)$ measured at the $D$ port (up to diminutive differences at transition frequencies corresponding to the bosonic isotopes of erbium), it sufficiently characterizes the Doppler broadened absorption profile of the total transmitted probe beam. We observe increased absorption of the probe beam for higher HCL currents. This is a direct consequence of increased atomic density in the probing region. The asymmetry of the absorption dip reflects the fact that absorption is stronger for the $^{166}$Er transition than for the $^{168}$Er and $^{170}$Er transitions. The increased absorption for higher HCL currents can be characterized in terms of optical density, OD~$=-\mathrm{ln} (T_{\mathrm{min}}/T_{\mathrm{max}})$, where $T_{\mathrm{min}}/T_{\mathrm{max}}$ is the ratio of the lowest to the highest transmitted probe signal values detected over the range of frequency values explored.  As shown in Fig. \pref{HCL-plot}{(d)} the optical density increases roughly linearly over the range of investigated HCL current values. For the stated conditions on the pump and probe beam powers and sizes, it is advisable to operate at lower HCL currents to both increase the signal slope and to extend the operational lifetime of the HCL, depending on the spectral feature chosen to use for laser stabilization.

\subsection{Pump laser polarization}
\label{pumppolarization}
\begin{figure}[t!]
	\includegraphics[width= \columnwidth]{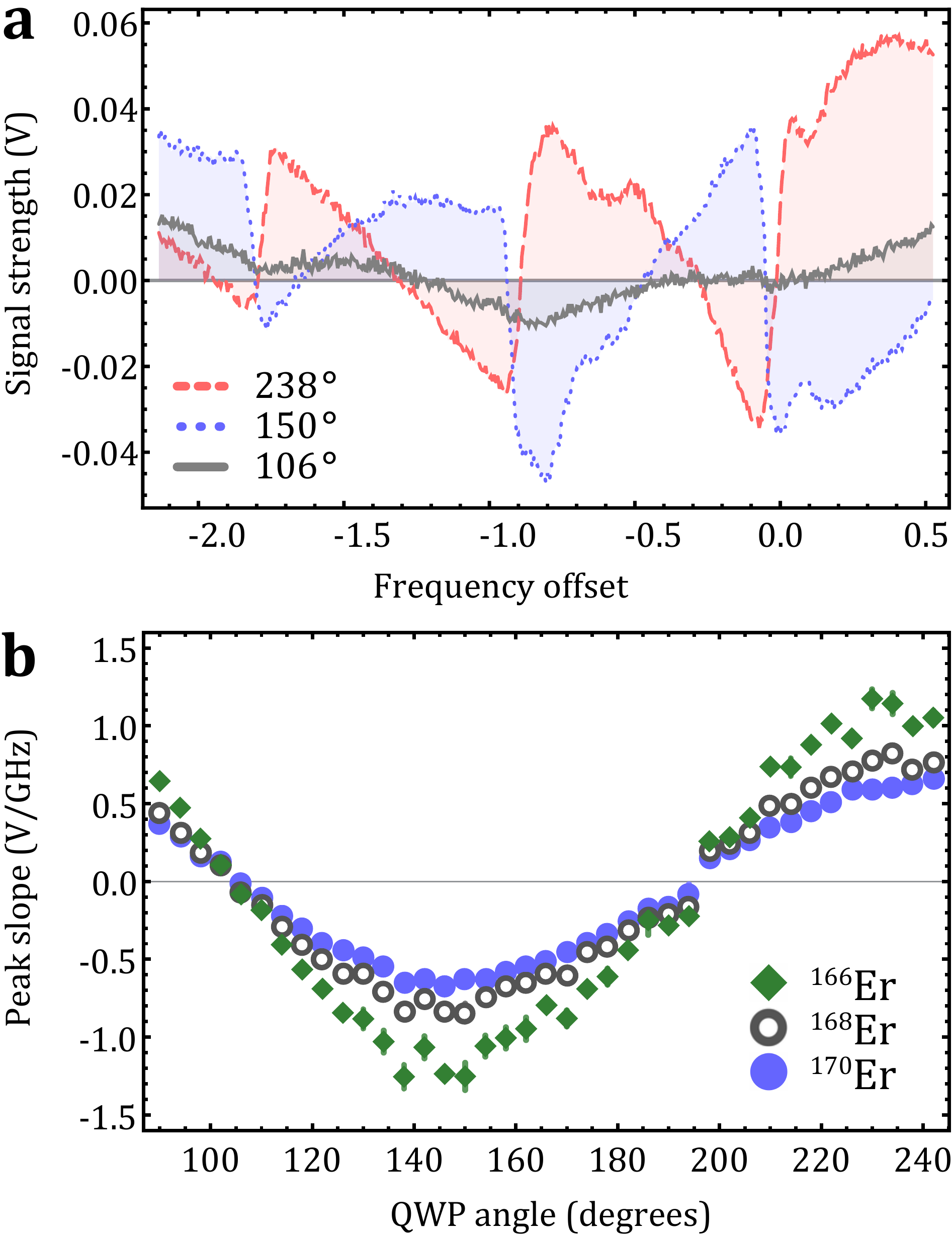}
	\caption{Dependence of polarization spectra on the pump laser polarization. 
		(a)~Polarization spectra for the cases of a vertically-polarized pump beam, with QWP angle of \squig 106 degrees (gray solid line), a right-circularly-polarized pump beam, with a QWP angle of \squig 238 degrees (red dashed line), and a left-circularly-polarized pump beam, with a QWP angle of \squig 150 degrees (blue dotted line).
		(b)~Peak slope as a function of the QWP angle for the three main spectral features relating to the bosonic isotopes $^{166}$Er, $^{168}$Er, and $^{170}$Er. All measurements in this figure were made for a pump peak intensity of 7.9(1)~mW/mm$^2$, a probe peak intensity of 0.536(5)~mW/mm$^2$, and an applied HCL current of 5.5~mA. Error bars denote one standard error.
	}
	\label{Polar-plot}
\end{figure}
The magnitude of the dispersive signal obtained from the polarization spectroscopy setup is directly proportional to the amount of rotation of the polarization of the probe beam when it passes through the atomic medium. This rotation, induced by anisotropic population distribution in the atomic medium by the pump beam, is controlled by the QWP setting.  Tuning the polarization of the pump beam from right-circular to left-circular polarization should lead to a change in the sign of the slopes of the dispersive signal from positive to negative, with a vanishing slope for linear polarization.
Figure~\pref{Polar-plot}{(a)} shows three different spectra relating to right-circular (red dashed line), left-circular (blue dotted line), and linear (grey solid line) pump beam polarizations. Similar experimental parameters as in Fig.~\pref{HCL-plot}{(a)} are used: a pump peak intensity of 7.9(1)~mW/mm$^2$, a probe peak intensity of 0.535(5)~mW/mm$^2$, and an applied HCL current of 5.5~mA. It can readily be seen that as the polarization switches from right-circular to left-circular, the slopes of the dispersive features in the polarization spectra switch sign. This is due to an inversion in the sign of circular dichroism and rotational birefringence induced in the atomic medium, as the change in pump beam polarization leads to a change in the induced anisotropy of the internal state population distribution. For the case of a linearly-polarized pump beam, only the smoothly-varying background of the polarization spectroscopy signal (which is not completely background free, unlike for the case of related modulation-based techniques~\cite{frisch}) is observed, along with diminutive minima at transition frequencies corresponding to the bosonic isotopes of erbium.

For the three strongest features related to the bosonic isotopes $^{166}$Er, $^{168}$Er, and $^{170}$Er, we plot in Fig.~\pref{Polar-plot}{(b)} the fit-determined signal slopes as a function of the pump beam polarization (set by the QWP angle). For a QWP angle of \squig 106 degrees (which corresponds to the fast axis being aligned with the incident vertical polarization), the pump beam remains vertically-polarized, and a near zero slope is found for the different features. A similarly vanishing slope is found for an angle of \squig 195 degrees. Overall, the three different signal slopes are seen to vary in a sinusoidal fashion, peaking in magnitude when the pump beam is either right-circularly- (angle of \squig 238 degrees) or left-circularly- (angle of \squig 150 degrees) polarized. This behavior is in good agreement with the expected dependence on the induced population anisotropy in the pumped atomic medium.

\subsection{Pump and probe peak intensity}
\label{pumpprobeintensity}

We examine in this section the dependence of the signal slope of dispersive signals associated with the $^{166}$Er, $^{168}$Er, and $^{170}$Er isotopes on pump and the probe beam intensities. An increase in the right-circularly-polarized pump beam peak intensity leads to optical pumping of the atoms to states with larger magnetic quantum numbers, thus causing a corresponding increase in the rotational birefringence and circular dichroism associated with the anisotropic distribution of population. This leads to larger rotations of the polarization of the probe beam and corresponds to an increased magnitude of the dispersive signal for low values of the saturation parameter $s=I_{\mathrm{peak}}/I_{\mathrm{sat}}$ (ratio of the pump beam peak intensity to the saturation intensity, where $I_{\mathrm{sat}}\approx 0.6~\mathrm{mW/mm}^2$ for the 401-nm transition of erbium). When $s>1$, there should be diminished increase in population anisotropy since the atoms have already been pumped to closed transitions. For increasing pump beam peak intensities, a plateau should develop in the dependence of the signal slope on $s$.
\begin{figure*}[t!]
	\includegraphics[width=2\columnwidth]{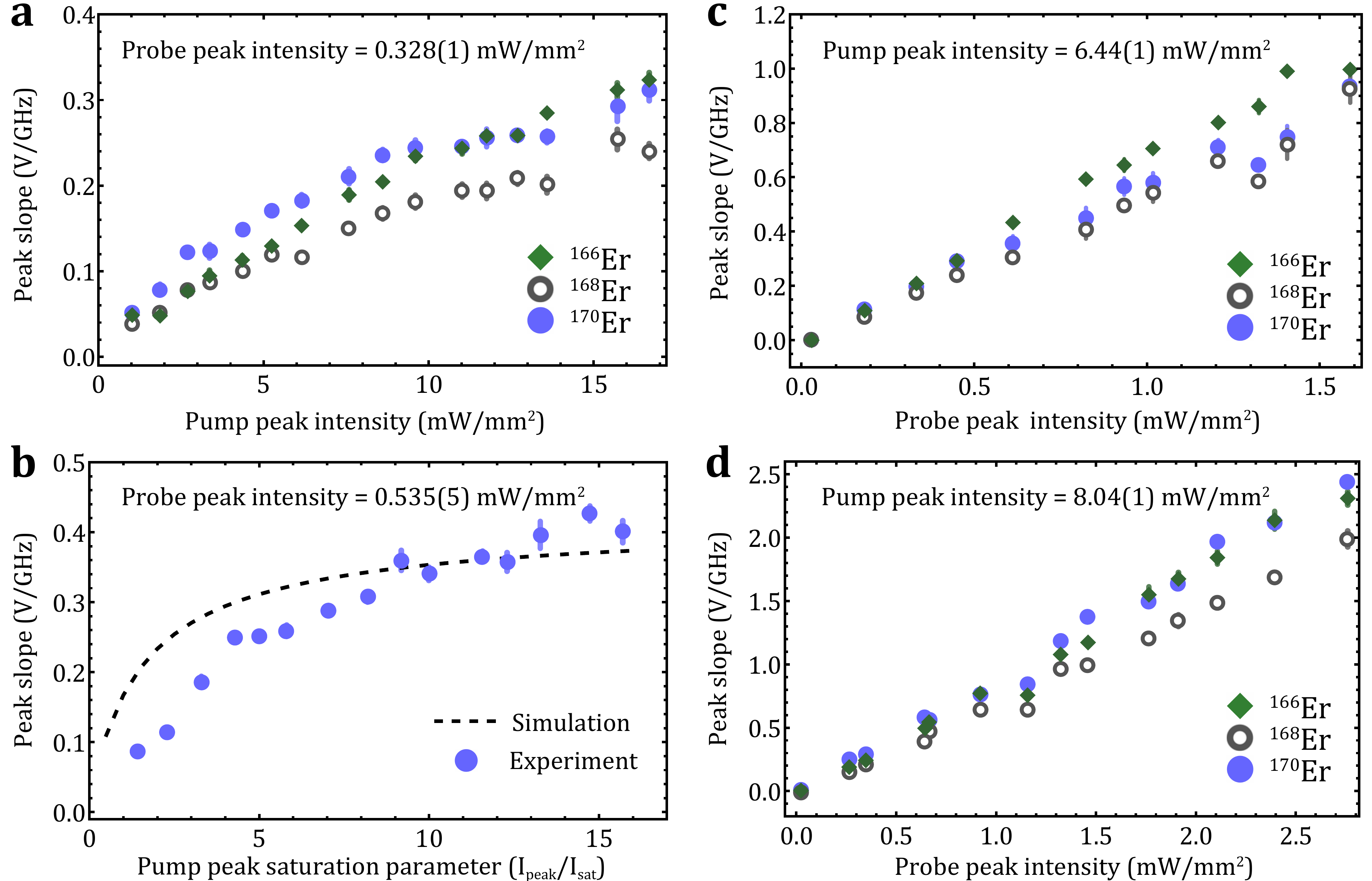}
	\caption{Dependence on pump and probe beam intensities. (a)~Peak slope of the dispersive signal for $^{166}$Er, $^{168}$Er, and $^{170}$Er with the probe beam fixed at 0.328(1) mW/mm$^2$ while varying the peak intensity of the pump beam. 
		(b) Fitted simulation of the dependence of the peak slope on the peak intensity of the pump beam. The data is plotted for peak intensity of the pump beam for $^{170}$Er with the probe beam fixed at 0.535(5) mW/mm$^2$. The HCL current is set at 7.5 mA for these measurements. (c)~Peak slope of the dispersive signals for $^{166}$Er, $^{168}$Er, and $^{170}$Er, as a function of the probe beam peak intensity for a fixed pump peak intensity of 6.44(1)~mW/mm$^2$. At this HCL value, the peak slopes are not correlated with the natural abundances of the three bosonic isotopes, as explained in Section \pref{results}{-A}. (d)~Peak slope of the dispersive signals for $^{166}$Er, $^{168}$Er, and $^{170}$Er with the pump beam peak intensity set at 8.04(1)~mW/mm$^2$ and varying the probe beam peak intensity. The magnitude of the peak slope increases linearly with the peak intensity of probe beam used. These measurements were taken at an HCL current value 7.5 mA. Error bars denote one standard error.}	
	\label{pumpdep}
\end{figure*}

Figure \pref{pumpdep}{(a)} shows the measured dependence of the slope of the dispersive signal on the pump beam peak intensity expressed in terms of $s$. The probe beam peak intensity was fixed at 0.328(1) mW/mm$^2$ while the peak intensity of the pump beam was varied between 1.02(2)~mW/mm$^2$ and 16.7(1)~mW/mm$^2$. The signal slope increases rapidly for low peak intensity values of the pump beam and then levels off for higher intensities. Since we are operating at an HCL current of 7.5 mA, we are in a regime where the signal slope associated with the bosonic isotopes do not correlate with the natural abundances, but are rather nearly equal. This is described in more detail in Fig. \pref{HCL-plot}{(a)}.

Following Section \pref{theory}{-A} we numerically simulate the dependence of the dispersive signal on the peak saturation parameter associated with the pump beam. As shown in Fig.~\pref{pumpdep}{(b)} we observe a rapid increase in signal slope that slows down for higher values of saturation parameter. The theoretical curve is compared with a polarization spectroscopy signal measured for a probe peak intensity of 0.535(5) mW/mm$^2$. We find reasonably good agreement with the data taken.

The theoretical curve accounts for the inhomogeneous spatial distribution of the intensity of the pump beam in the area of interrogation using
\begin{equation}
f'(s_\mathrm{peak})=\frac{\int_{0}^{\infty} f(s(I(r)))~I(r) dr}{\int_{0}^{\infty}I(r)dr},
\end{equation}
where $f(s)= \mathcal{A}(s, J, J') l'$ is proportional to the slope of the dispersive signal. The time averaged anisotropy, $\mathcal{A}(s, J, J')$, discussed in more detail in Section \ref{theory}, relates to the simulated peak height of the dispersive signal for a transition $(J,J')=(6,7)$ for the bosonic isotopes. The effective length of the region of overlap between the pump and the probe beam is given by $l'$. The saturation parameter, $s$, is defined as a function of a spatially varying intensity characteristic of a Gaussian beam, i.e. $I(r) =I_{\mathrm{peak}}\mathrm{Exp}(-2r^2/\sigma^2)$, with the $1/e^2$ half-width given by $\sigma$.

To determine the dependence of the dispersive signal on the probe peak intensity, we fixed the pump peak intensity at  6.44(1) mW/mm$^2$ and 8.04(1) mW/mm$^2$ in Fig. \pref{pumpdep}{(c)} and \pref{pumpdep}{(d)}, respectively. The HCL current in this case was 7.5 mA. As we increased the probe beam peak intensity we observed an increase in the magnitude of the dispersive features associated with the three most abundant bosonic isotopes ($^{166}$Er, $^{168}$Er, and $^{170}$Er). This is due to the increased flux of photons in the region of interrogation, which in turn leads to an increased photon count at the balanced detector and an increase in the magnitude of the dispersive signal. A roughly linear dependence on the probe peak intensity is observed for all the features.

\section{Theoretical model}
\label{theory}
\subsection{Polarization spectroscopy signal modeling}
\label{theory1}
We describe a general approach used to simulate polarization spectra by solving rate equations for all possible transitions that can be induced by the right-circularly-polarized pump beam resonant with the $^3H_6 \rightarrow  ^1P_1$ transition in erbium.
As shown in more detail in Ref.~\cite{pearman}, the difference signal for a single transition is given by
\begin{equation}
S_{\mathrm{diff}}(\omega) \propto \mathrm{e}^{- \bar{\alpha} l } \bigg ( \Delta \alpha ~ l' \frac{x(\omega)}{1+{x(\omega)}^2} \bigg ) ,
\label{signal}
\end{equation}
where $x(\omega)= \frac{\omega-\omega_0}{\Gamma/2}$,  $\Gamma$ is the spontaneous decay rate of the atom ($\Gamma = 2 \pi \times 29.7~ \mathrm{ MHz} $ for the 401 nm line of erbium \cite{hartog,mclelland2,lawler}), $l$ is the length of the active area of interrogation (section of the HCL occupied by the sputtered erbium atoms), $l'$ is the effective length of interrogation where the pump and probe beam overlap, $\bar{\alpha}$ is the average of the absorption coefficients corresponding to right- and left-circular polarizations pump beams, and $\Delta \alpha$ is the anisotropy, i.e. the difference between these two absorption coefficients. 

The pump beam induced anisotropy is given by

\begin{equation}
\begin{split}
&\Delta \alpha(F, F', t) =\\
&n \sigma_0\sum_{m_F=-F}^F \mathcal{R}_{(F, m_F) \rightarrow (F',m_F+1)} (\rho_{F, m_F}-\rho'_{F', m_F+1}) \\
&\quad \quad \quad \quad \quad - \mathcal{R}_{(F, m_F) \rightarrow (F', m_F-1)} (\rho_{F, m_F}-\rho'_{F', m_F-1}),
 \end{split}
 \label{timedani}
\end{equation}
where $n$ and $\sigma_0$ are the atomic density and resonant absorption cross-section, respectively \cite{Do,demtroederv2,harris}.~The initial and final hyperfine states are labeled by $F$ and $F'$. Since $I=0$ for the bosonic isotopes, we consider $J=6 \rightarrow J'=7$ transitions. In general, we consider transitions that involve the $^3H_6$ ground state and the $^1P_1$ excited state. The transition line strength,
$\mathcal{R}_{(F, m_F) \rightarrow (F', m_F')}$, is given by the square magnitude of the dipole matrix element for the transition between two levels~\cite{metcalf,Yoshikawa:03}, i.e.
\begin{equation}
\begin{split}
& \mathcal{R}_{(F, m_F) \rightarrow (F', m_F')} = |e \braket{n' L' ||r|| n L}|^2 \\ &\times (2 J' +1) (2 J +1) (2 F' +1) (2 F +1)\\
& \times \bigg [ \bigg \{ \begin{array}{ccc}
 L' & J' & S \\
 J & L & 1 \\
\end{array}  \bigg \}
\bigg \{ \begin{array}{ccc}
 J' & F' & I \\
 F & J & 1 \\
\end{array}  \bigg \}
\bigg ( \begin{array}{ccc}
 F & 1 & F' \\
m_{F} & \Delta m & - m_{F'}\\
\end{array}  \bigg )
\bigg]^2,
 \end{split}
\end{equation}
where the quantity $\braket{\mathrm{final} ||r||\mathrm{initial}}$ is the reduced matrix element which can be simplified using the Wigner-Eckart theorem~\cite{Wybourne}, $L$ and $S$ are the electronic orbital and spin angular momentum quantum numbers, respectively, while $I$ is the nuclear spin angular momentum. The hyperfine quantum number, $F$, ranges from $F=|I-J|$ to $F=|I+J|$, while the projection of $F$ onto the $z$-axis, $m_F$, is the hyperfine sub-level.
The factors $\{ \dots\}$ and $( \dots )$ are the 6J symbol and 3J symbol, respectively, and can be computed using symbolic software packages. The fractional population of the $m_F$ sublevel is labeled as $\rho_{F,m_F}$.

The population dynamics are therefore described by the rate equations generalized for multilevel atoms \cite{harris}, i.e.
\begin{equation}
\begin{split}
\frac{d \rho_{F, m_F}}{d t}
  = - \sum_{F'=F-1}^{F'=F+1} \mathcal{R}_{(F, m_F) \rightarrow (F', m_F+\Delta m)} \frac{\Gamma}{2}s\\ \times \frac{(\rho_{F, m_F}-\rho'_{F', m_F+\Delta m})}{1+4 (\tilde{\Delta}/\Gamma)^2} \\
 - \sum_{m_{F'}=m_{F-1}}^{m_{F'}=m_{F+1}} \sum_{F'=F-1}^{F'=F+1}
 \mathcal{R}_{(F, m_F) \rightarrow (F', m_{F'})} \Gamma \rho'_{F', m_{F'}}.
 \end{split}
\end{equation}
For the change in the hyperfine sub-level, $\Delta m=\pm 1$ represents $\sigma^{\pm}$ transitions while $\Delta m=0$ represents $\pi$ transitions. The saturation parameter is given by $s=I/I_{\mathrm{sat}}$ where $I$ is the beam intensity and $I_{\mathrm{sat}}$ is the saturation intensity. The detuning associated with the transition $F \rightarrow F'$ is given by $\tilde{\Delta}= \Delta + \Delta_{F,F'}$ where $ \Delta_{F,F'}$ is the detuning from a chosen transition (the transition associated with $^{166}$Er isotope in our case). The last term corresponds to spontaneous emission out of the excited state.

The difference signal measured in experiment is proportional to a time averaged anisotropy. Following the discussion in Ref.~\cite{harris}, we account for this in the model by introducing a weighting function that characterizes the atomic trajectories transverse to the direction of propagation of the beam. These trajectories correspond to a class of atoms with a vanishing velocity component along the beam. The weighting function, given by
\begin{equation}
\begin{split}
W(t)=\int_0^{2 a}f(z)~ g(z,t) dz,
\end{split}
\end{equation}
consists of a probability distribution $f(z)$ for different trajectory path lengths $z \in (0,2a)$ where $a$ is the radius of the beam. In our case we consider uniform probabilities for all path lengths, i.e. $f(z)=1/2a$. The second component of the weighting function relates to the distribution, $g(z,t)$, of interrogation times for the trajectories of path length $z$ transverse to the beam. Even though the distribution of sputtered erbium atoms in the HCL is not necessarily thermal, the velocity profile of atoms can be effectively captured by a Maxwell-Boltzmann distribution. The expression becomes
\begin{equation}
\begin{split}
W(t) = \int_0^{2 a} \frac{1}{2 a} \frac{m z^2}{k_BT t^2} \mathrm{Exp} \bigg ( -\frac{m z^2}{2 k_BT t^2}\bigg) dz.
\end{split}
\end{equation}
The mean velocity associated with $g(z,t)$ can be  estimated from the average temperature of the sputtered atoms produced at the HCL.
By fitting three convoluted Gaussian distributions, centered at the transition frequencies corresponding to the three dominant bosonic isotopes, to the Doppler broadened signal obtained from the A port of the balanced detector, we estimate the temperature of the sputtered atoms to be $\squig 1000-1200$~K. This corresponds to velocity values $\squig 300-350$~m/s.

The time averaged anisotropy is therefore given by \cite{harris}
\begin{equation}
\mathcal{A} (F,F')=\int_{0}^{t_{\mathrm{final}}}\frac{1}{t_{\mathrm{final}}} \Delta \alpha (F,F',t)
W(t)  dt,
\label{timeavgani}
\end{equation}
where $t_\mathrm{final}$ is the total interrogation time. Considering the characteristic velocity ($\squig 3 \times 10^5$~mm/s) and the size of the pump beam ($\squig 1$~mm), the dynamics have essentially ceased in a few microseconds. By choosing $t_\mathrm{final}=10~\mu$s we take, effectively, the long time limit. Since $I=0$ for bosonic isotopes, we refer to the time-averaged anisotropy at the bosonic resonances as $\mathcal{A}(J,J')$ in Section \pref{results}{-C}.

While $\mathcal{A} (F,F')$ only depends on the pump beam parameters, the attenuation of the probe signal, governed by the Beer-Lambert law, is captured in the pre-factor 
\begin{equation}
 e^{-\bar{\alpha} l} ~\squig~ T(\omega)/T_{\mathrm{max}},
\end{equation}
where $T(\omega)/T_{\mathrm{max}}$ is the transmitted probe signal, normalized to the incident probe signal obtained from the $A$ output of the photodetector. That is, we convolute the numerically simulated lineshape with $T(\omega)/T_{\mathrm{max}}$, the observed (normalized) absorption profile.
Using Eq.~\eqref{timedani} with $\Delta \alpha \rightarrow \mathcal{A}(F, F')$, we can now model the expected lineshapes. Crossover peaks between two features are calculated by taking the mean $\mathcal{A} (F, F')$ of the associated transitions~\cite{Yoshikawa:03}. The simulated polarization spectroscopy lineshape is therefore given by
\begin{equation}
S_{\mathrm{diff}}(\omega) \propto \sum_{\{F, F'\}} \frac{T(\omega)}{T_{\mathrm{max}}} \bigg(\mathcal{A}(F,F')~l' \frac{x(\omega)}{1+{x(\omega)}^2}\bigg).
\label{signal2}
\end{equation}
The simulation assumes near-resonant pump beam frequency for each transition and adds up the dispersive signals relating to all the transitions to construct the final lineshape.

\subsection{Lineshape fitting and analysis}
\label{lineshpfitting}

\begin{figure}[t!]
	\includegraphics[width=\columnwidth]{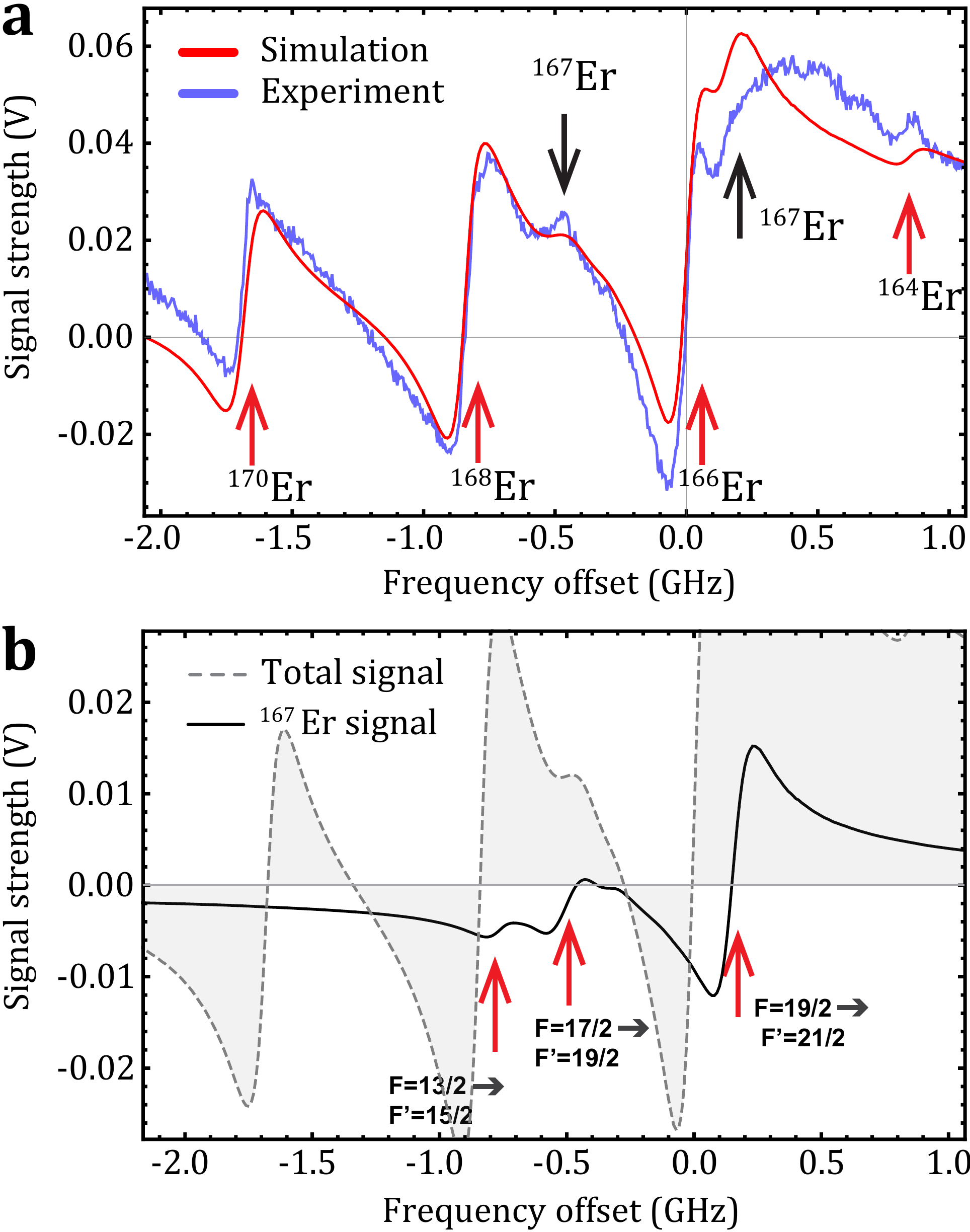}
	\caption{Lineshape modeling.  (a)~Theoretically calculated fit using equations presented in Section \pref{theory}{-A} overlaid on the measured trace. The linewidth used in theory is extracted by fitting the second derivative of a Lorentzian to the discrete derivative of the experimental lineshape for the bosonic isotopes $^{166}$Er, $^{168}$Er, and $^{170}$Er and taking an average. (b)~ Dispersive signal corresponding to fermionic hyperfine transitions obtained from rate calculations. The features corresponding to $F=13/2 \rightarrow F'=15/2$, $F=17/2 \rightarrow F'=19/2$ and $F=19/2 \rightarrow F'=21/2$ have the largest magnitudes. Signals corresponding to bosonic transitions have been left out for clarity. The dashed gray line shows the predicted lineshape that includes both the bosonic and fermionic features. }
	\label{fermionicAni}
\end{figure}

Using the results from Section \pref{theory}{-A}, we model the lineshape produced by polarization spectroscopy for the case of erbium. We consider a right-circularly-polarized pump beam which induces transitions for which the hyperfine sub-level changes by $\Delta m_F=+1$. 

Figure~\pref{fermionicAni}{(a)} compares the experimental polarization spectra to our theoretical model. The lineshape (calculated using Eq. \eqref{signal2}) considers the experimental conditions used in Fig. \pref{lorentzianfit}{(a)}, i.e. a peak pump intensity of 7.9(1)~mW/mm$^2$ and $1/e^2$ beam radius of $0.447(1)$~mm (average of the $1/e^2$ widths of the horizontal and vertical axis of the beam). The experimental lineshape is taken at an HCL current of 5.5~mA and probe peak intensity of 0.535(5)~mW/mm$^2$. The experimental and theoretical spectra show three dominant features, relating to transitions associated with the bosonic isotopes $^{166}$Er, $^{168}$Er, and $^{170}$Er. Some additional small features corresponding to fermionic isotope $^{167}$Er and the bosonic isotope $^{164}$Er can be seen as well. 

The linewidth used in the theoretical model is obtained by fitting the second derivative of a Lorentzian to the discrete derivative of dispersive features that correspond to the three most abundant bosonic isotopes. We extract an effective linewidth of $\Gamma_\mathrm{eff}/2\pi=145(17)$~MHz. This is substantially higher (by a factor of \squig 5) than the natural linewidth ($\Gamma/2\pi = 29.7(6)$~MHz~\cite{frisch}). We attribute the broadening, in part, to power broadening, contributing a factor of $\sqrt{1+I_{\mathrm{peak}}/I_{\mathrm{sat}}}= 3.77$ for a pump peak intensity of $7.9$~mW/mm$^2$. Collisional broadening and residual Doppler-broadening due to a slight misalignment of the pump and probe beams can also contribute to the observed linewidth.

Due to such broadening and the relatively close spacing of the fermionic and bosonic resonances (on the order of the broadened linewidth) the fermionic features are mostly obscured. However, two features relating to the fermionic isotope $^{167}$Er can be clearly distinguished as shown in Fig. \pref{fermionicAni}{(b)}. One feature, appearing at a detuning of roughly 0.5 GHz below the $^{166}$Er resonance, relates to transitions involving the states $(F,m_F,F',m_{F'}) = (17/2, 17/2, 19/2, 19/2)$. Another feature involving the states $(F,m_F,F',m_{F'}) = (19/2, 19/2, 21/2, 21/2)$ appears as a peak and shoulder at frequencies just above ($\sim 0.15$ GHz) the $^{166}$Er resonance. The feature related to the states $(F,m_F,F',m_{F'}) = (13/2, 13/2, 15/2, 15/2)$ is much smaller than the nearby $^{168}$Er feature and is therefore completely obscured.

Up to an overall proportionality factor, we observe fairly good agreement between the numerically simulated lineshape and the measured lineshape despite the fact that polarization spectroscopy includes a Doppler-broadened background. While the fermionic transitions are mostly blurred out, the three most abundant bosonic isotope offer clear dispersive signals that are adequately spaced (\squig 1 GHz apart). These can readily be used as error signals for laser stabilization.

\section{Conclusions}
\label{conclusions}

We have presented a comparison between theory and experiment on the polarization spectroscopy of atomic erbium. We present simulations of the polarization spectra for the 401-nm transitions based on calculation of the population anisotropy in the atomic medium induced by a circularly-polarized pump laser beam, relating to an induced dichroism and birefringence. We find good qualitative agreement between the predicted spectra and the measured experimental lineshapes. The observed stability, signal strengths, and linewidths suggest that the simple method of polarization spectroscopy is suitable for laser stabilization on the near-UV transitions of the lanthanides. We analyze spectroscopic lineshapes for transitions relating to the various bosonic isotopes, and find that the signal slope grows linearly with the probe beam peak intensity. As a function of the pump beam peak intensity, we measured an initial large growth in the signal followed by saturation (as the induced level anisotropy saturates). The determination of ideal spectroscopy parameters for laser stabilization allows us to stabilize our diode laser for laser cooling and trapping atomic erbium in future experiments.

\appendix
\section{Fermionic erbium hyperfine transitions}
Table \ref{detunings} shows $^{167}$Er hyperfine transitions organized by detuning from $^{166}$Er. The detuning values are calculated using hyperfine constants $A_e/h =-100$ MHz and $B_e/h =-3079$ MHz for the $^1P_1$ excited state and  $A_g/h =-120$ MHz and $B_g/h =-4552$ MHz for the $^3H_6$ ground state. The frequency separation between $^{167}$Er and $^{166}$Er is taken to be -297~MHz. These values are taken from Ref.~\cite{frisch}.

\begin{table}
	\begin{tabular}{c|c|c|c}
		\hline \hline
		Transition	& Detuning &Transition  &Detuning \\ 
		$F \rightarrow F'$&(MHz)&$F \rightarrow F'$ &(MHz)\\
		\hline
		$15/2 \rightarrow 17/2$& -764 &$13/2 \rightarrow 11/2$& -87\\ 
		
		$13/2 \rightarrow 15/2$	& -760 &$7/2 \rightarrow 9/2$  & -37\\ 
		
		$11/2 \rightarrow 13/2$	&-584  & $15/2 \rightarrow 15/2$ & 106\\ 
		
		$17/2 \rightarrow 19/2$	& -499 &$19/2 \rightarrow 21/2$ & 150\\ 
		
		$9/2 \rightarrow 7/2$	& -420 &$5/2 \rightarrow 7/2$ & 213\\ 
		
		$11/2 \rightarrow 11/2$	& -390 & $15/2 \rightarrow 13/2$ & 585 \\ 
		
		$11/2 \rightarrow 9/2$	& -384  &  $17/2 \rightarrow 17/2$  &888 \\ 
		
		$9/2 \rightarrow 11/2$	& -320 & $17/2 \rightarrow 15/2$  &1758 \\ 
		
		$9/2 \rightarrow 9/2$	& -315 & $19/2 \rightarrow 19/2$  & 2190 \\ 
		
		$13/2 \rightarrow 13/2$& -282  &$19/2 \rightarrow 17/2$  & 3576 \\ 
		
		$7/2 \rightarrow 7/2$& -142 &  &  \\ 
		\hline \hline
	\end{tabular} 
	\caption {$^{167}$Er hyperfine transitions organized by detuning from $^{166}$Er transition between the $^3H_6$ ground state and the $^1P_1$ excited state.} \label{detunings}
\end{table}

\section*{Acknowledgements}
We thank Fangzhao Alex An for a critical reading of the manuscript. This work was supported by the Office of Naval Research Science and Technology (ONR; Award N00014-16-1-2895).

%\bibliographystyle{apsrev4-1}
%\bibliographystyle{unsrtnat}
%\bibliography{ErSpectro}

\begin{thebibliography}{32}%
	\makeatletter
	\providecommand \@ifxundefined [1]{%
		\@ifx{#1\undefined}
	}%
	\providecommand \@ifnum [1]{%
		\ifnum #1\expandafter \@firstoftwo
		\else \expandafter \@secondoftwo
		\fi
	}%
	\providecommand \@ifx [1]{%
		\ifx #1\expandafter \@firstoftwo
		\else \expandafter \@secondoftwo
		\fi
	}%
	\providecommand \natexlab [1]{#1}%
	\providecommand \enquote  [1]{``#1''}%
	\providecommand \bibnamefont  [1]{#1}%
	\providecommand \bibfnamefont [1]{#1}%
	\providecommand \citenamefont [1]{#1}%
	\providecommand \href@noop [0]{\@secondoftwo}%
	\providecommand \href [0]{\begingroup \@sanitize@url \@href}%
	\providecommand \@href[1]{\@@startlink{#1}\@@href}%
	\providecommand \@@href[1]{\endgroup#1\@@endlink}%
	\providecommand \@sanitize@url [0]{\catcode `\\12\catcode `\$12\catcode
		`\&12\catcode `\#12\catcode `\^12\catcode `\_12\catcode `\%12\relax}%
	\providecommand \@@startlink[1]{}%
	\providecommand \@@endlink[0]{}%
	\providecommand \url  [0]{\begingroup\@sanitize@url \@url }%
	\providecommand \@url [1]{\endgroup\@href {#1}{\urlprefix }}%
	\providecommand \urlprefix  [0]{URL }%
	\providecommand \Eprint [0]{\href }%
	\providecommand \doibase [0]{http://dx.doi.org/}%
	\providecommand \selectlanguage [0]{\@gobble}%
	\providecommand \bibinfo  [0]{\@secondoftwo}%
	\providecommand \bibfield  [0]{\@secondoftwo}%
	\providecommand \translation [1]{[#1]}%
	\providecommand \BibitemOpen [0]{}%
	\providecommand \bibitemStop [0]{}%
	\providecommand \bibitemNoStop [0]{.\EOS\space}%
	\providecommand \EOS [0]{\spacefactor3000\relax}%
	\providecommand \BibitemShut  [1]{\csname bibitem#1\endcsname}%
	\let\auto@bib@innerbib\@empty
	%</preamble>
	\bibitem [{\citenamefont {Bloch}\ \emph {et~al.}(2008)\citenamefont {Bloch},
		\citenamefont {Dalibard},\ and\ \citenamefont {Zwerger}}]{bloch}%
	\BibitemOpen
	\bibfield  {author} {\bibinfo {author} {\bibfnamefont {I.}~\bibnamefont
			{Bloch}}, \bibinfo {author} {\bibfnamefont {J.}~\bibnamefont {Dalibard}}, \
		and\ \bibinfo {author} {\bibfnamefont {W.}~\bibnamefont {Zwerger}},\ }\href
	{\doibase 10.1103/RevModPhys.80.885} {\bibfield  {journal} {\bibinfo
			{journal} {Rev. Mod. Phys.}\ }\textbf {\bibinfo {volume} {80}},\ \bibinfo
		{pages} {885} (\bibinfo {year} {2008})}\BibitemShut {NoStop}%
	\bibitem [{\citenamefont {Lahaye}\ \emph {et~al.}(2009)\citenamefont {Lahaye},
		\citenamefont {Menotti}, \citenamefont {Santos}, \citenamefont {Lewenstein},\
		and\ \citenamefont {Pfau}}]{lahaye:dipolar_review_2009}%
	\BibitemOpen
	\bibfield  {author} {\bibinfo {author} {\bibfnamefont {T.}~\bibnamefont
			{Lahaye}}, \bibinfo {author} {\bibfnamefont {C.}~\bibnamefont {Menotti}},
		\bibinfo {author} {\bibfnamefont {L.}~\bibnamefont {Santos}}, \bibinfo
		{author} {\bibfnamefont {M.}~\bibnamefont {Lewenstein}}, \ and\ \bibinfo
		{author} {\bibfnamefont {T.}~\bibnamefont {Pfau}},\ }\href {\doibase
		10.1088/0034-4885/72/12/126401} {\bibfield  {journal} {\bibinfo  {journal}
			{Rep. Prog. Phys.}\ }\textbf {\bibinfo {volume} {72}},\ \bibinfo {pages}
		{126401} (\bibinfo {year} {2009})}\BibitemShut {NoStop}%
	\bibitem [{\citenamefont {Collaboration}\ \emph {et~al.}(2014)\citenamefont
		{Collaboration}, \citenamefont {Baron}, \citenamefont {Campbell},
		\citenamefont {DeMille}, \citenamefont {Doyle}, \citenamefont {Gabrielse},
		\citenamefont {Gurevich}, \citenamefont {Hess}, \citenamefont {Hutzler},
		\citenamefont {Kirilov}, \citenamefont {Kozyryev}, \citenamefont
		{O¡¯Leary}, \citenamefont {Panda}, \citenamefont {Parsons}, \citenamefont
		{Petrik}, \citenamefont {Spaun}, \citenamefont {Vutha},\ and\ \citenamefont
		{West}}]{ACME:EDM2014}%
	\BibitemOpen
	\bibfield  {author} {\bibinfo {author} {\bibfnamefont {The ACME}\ \bibnamefont
			{Collaboration}}, \bibinfo {author} {\bibfnamefont {J.}~\bibnamefont
			{Baron}}, \bibinfo {author} {\bibfnamefont {W.~C.}\ \bibnamefont {Campbell}},
		\bibinfo {author} {\bibfnamefont {D.}~\bibnamefont {DeMille}}, \bibinfo
		{author} {\bibfnamefont {J.~M.}\ \bibnamefont {Doyle}}, \bibinfo {author}
		{\bibfnamefont {G.}~\bibnamefont {Gabrielse}}, \bibinfo {author}
		{\bibfnamefont {Y.~V.}\ \bibnamefont {Gurevich}}, \bibinfo {author}
		{\bibfnamefont {P.~W.}\ \bibnamefont {Hess}}, \bibinfo {author}
		{\bibfnamefont {N.~R.}\ \bibnamefont {Hutzler}}, \bibinfo {author}
		{\bibfnamefont {E.}~\bibnamefont {Kirilov}}, \bibinfo {author} {\bibfnamefont
			{I.}~\bibnamefont {Kozyryev}}, \bibinfo {author} {\bibfnamefont {B.~R.}\
			\bibnamefont {O¡¯Leary}}, \bibinfo {author} {\bibfnamefont {C.~D.}\
			\bibnamefont {Panda}}, \bibinfo {author} {\bibfnamefont {M.~F.}\ \bibnamefont
			{Parsons}}, \bibinfo {author} {\bibfnamefont {E.~S.}\ \bibnamefont {Petrik}},
		\bibinfo {author} {\bibfnamefont {B.}~\bibnamefont {Spaun}}, \bibinfo
		{author} {\bibfnamefont {A.~C.}\ \bibnamefont {Vutha}}, \ and\ \bibinfo
		{author} {\bibfnamefont {A.~D.}\ \bibnamefont {West}},\ }\href {\doibase
		10.1126/science.1248213} {\bibfield  {journal} {\bibinfo  {journal}
			{Science}\ }\textbf {\bibinfo {volume} {343}},\ \bibinfo {pages} {269}
		(\bibinfo {year} {2014})}\BibitemShut {NoStop}%
	\bibitem [{\citenamefont {Saffman}\ and\ \citenamefont
		{M\o{}lmer}(2008)}]{HolmiumComputing}%
	\BibitemOpen
	\bibfield  {author} {\bibinfo {author} {\bibfnamefont {M.}~\bibnamefont
			{Saffman}}\ and\ \bibinfo {author} {\bibfnamefont {K.}~\bibnamefont
			{M\o{}lmer}},\ }\href {\doibase 10.1103/PhysRevA.78.012336} {\bibfield
		{journal} {\bibinfo  {journal} {Phys. Rev. A}\ }\textbf {\bibinfo {volume}
			{78}},\ \bibinfo {pages} {012336} (\bibinfo {year} {2008})}\BibitemShut
	{NoStop}%
	\bibitem [{\citenamefont {Bell}\ \emph {et~al.}(1999)\citenamefont {Bell},
		\citenamefont {Stuhler}, \citenamefont {Locher}, \citenamefont {Hensler},
		\citenamefont {Mlynek},\ and\ \citenamefont {Pfau}}]{ChromiumMOT}%
	\BibitemOpen
	\bibfield  {author} {\bibinfo {author} {\bibfnamefont {A.~S.}\ \bibnamefont
			{Bell}}, \bibinfo {author} {\bibfnamefont {J.}~\bibnamefont {Stuhler}},
		\bibinfo {author} {\bibfnamefont {S.}~\bibnamefont {Locher}}, \bibinfo
		{author} {\bibfnamefont {S.}~\bibnamefont {Hensler}}, \bibinfo {author}
		{\bibfnamefont {J.}~\bibnamefont {Mlynek}}, \ and\ \bibinfo {author}
		{\bibfnamefont {T.}~\bibnamefont {Pfau}},\ }\href
	{http://stacks.iop.org/0295-5075/45/i=2/a=156} {\bibfield  {journal}
		{\bibinfo  {journal} {EPL}\ }\textbf {\bibinfo {volume}
			{45}},\ \bibinfo {pages} {156} (\bibinfo {year} {1999})}\BibitemShut
	{NoStop}%
	\bibitem [{\citenamefont {Ban}\ \emph {et~al.}(2005)\citenamefont {Ban},
		\citenamefont {Jacka}, \citenamefont {Hanssen}, \citenamefont {Reader},\ and\
		\citenamefont {McClelland}}]{Ban:05}%
	\BibitemOpen
	\bibfield  {author} {\bibinfo {author} {\bibfnamefont {H.~Y.}\ \bibnamefont
			{Ban}}, \bibinfo {author} {\bibfnamefont {M.}~\bibnamefont {Jacka}}, \bibinfo
		{author} {\bibfnamefont {J.~L.}\ \bibnamefont {Hanssen}}, \bibinfo {author}
		{\bibfnamefont {J.}~\bibnamefont {Reader}}, \ and\ \bibinfo {author}
		{\bibfnamefont {J.~J.}\ \bibnamefont {McClelland}},\ }\href {\doibase
		10.1364/OPEX.13.003185} {\bibfield  {journal} {\bibinfo  {journal} {Opt.
				Express}\ }\textbf {\bibinfo {volume} {13}},\ \bibinfo {pages} {3185}
		(\bibinfo {year} {2005})}\BibitemShut {NoStop}%
	\bibitem [{\citenamefont {McClelland}(2006)}]{mclelland2}%
	\BibitemOpen
	\bibfield  {author} {\bibinfo {author} {\bibfnamefont {J.~J.}\ \bibnamefont
			{McClelland}},\ }\href {\doibase 10.1103/PhysRevA.73.064502} {\bibfield
		{journal} {\bibinfo  {journal} {Phys. Rev. A}\ }\textbf {\bibinfo {volume}
			{73}},\ \bibinfo {pages} {064502} (\bibinfo {year} {2006})}\BibitemShut
	{NoStop}%
	\bibitem [{\citenamefont {McClelland}\ and\ \citenamefont
		{Hanssen}(2006)}]{McClelland-Erb}%
	\BibitemOpen
	\bibfield  {author} {\bibinfo {author} {\bibfnamefont {J.~J.}\ \bibnamefont
			{McClelland}}\ and\ \bibinfo {author} {\bibfnamefont {J.~L.}\ \bibnamefont
			{Hanssen}},\ }\href {\doibase 10.1103/PhysRevLett.96.143005} {\bibfield
		{journal} {\bibinfo  {journal} {Phys. Rev. Lett.}\ }\textbf {\bibinfo
			{volume} {96}},\ \bibinfo {pages} {143005} (\bibinfo {year}
		{2006})}\BibitemShut {NoStop}%
	\bibitem [{\citenamefont {Griesmaier}\ \emph {et~al.}(2005)\citenamefont
		{Griesmaier}, \citenamefont {Werner}, \citenamefont {Hensler}, \citenamefont
		{Stuhler},\ and\ \citenamefont {Pfau}}]{GriesBEC}%
	\BibitemOpen
	\bibfield  {author} {\bibinfo {author} {\bibfnamefont {A.}~\bibnamefont
			{Griesmaier}}, \bibinfo {author} {\bibfnamefont {J.}~\bibnamefont {Werner}},
		\bibinfo {author} {\bibfnamefont {S.}~\bibnamefont {Hensler}}, \bibinfo
		{author} {\bibfnamefont {J.}~\bibnamefont {Stuhler}}, \ and\ \bibinfo
		{author} {\bibfnamefont {T.}~\bibnamefont {Pfau}},\ }\href {\doibase
		10.1103/PhysRevLett.94.160401} {\bibfield  {journal} {\bibinfo  {journal}
			{Phys. Rev. Lett.}\ }\textbf {\bibinfo {volume} {94}},\ \bibinfo {pages}
		{160401} (\bibinfo {year} {2005})}\BibitemShut {NoStop}%
	\bibitem [{\citenamefont {Naylor}\ \emph {et~al.}(2015)\citenamefont {Naylor},
		\citenamefont {Reigue}, \citenamefont {Mar\'echal}, \citenamefont {Gorceix},
		\citenamefont {Laburthe-Tolra},\ and\ \citenamefont {Vernac}}]{Cr-Fermi}%
	\BibitemOpen
	\bibfield  {author} {\bibinfo {author} {\bibfnamefont {B.}~\bibnamefont
			{Naylor}}, \bibinfo {author} {\bibfnamefont {A.}~\bibnamefont {Reigue}},
		\bibinfo {author} {\bibfnamefont {E.}~\bibnamefont {Mar\'echal}}, \bibinfo
		{author} {\bibfnamefont {O.}~\bibnamefont {Gorceix}}, \bibinfo {author}
		{\bibfnamefont {B.}~\bibnamefont {Laburthe-Tolra}}, \ and\ \bibinfo {author}
		{\bibfnamefont {L.}~\bibnamefont {Vernac}},\ }\href {\doibase
		10.1103/PhysRevA.91.011603} {\bibfield  {journal} {\bibinfo  {journal} {Phys.
				Rev. A}\ }\textbf {\bibinfo {volume} {91}},\ \bibinfo {pages} {011603}
		(\bibinfo {year} {2015})}\BibitemShut {NoStop}%
	\bibitem [{\citenamefont {Lu}\ \emph {et~al.}(2011)\citenamefont {Lu},
		\citenamefont {Burdick}, \citenamefont {Youn},\ and\ \citenamefont
		{Lev}}]{lu:strongly_Dy_2011}%
	\BibitemOpen
	\bibfield  {author} {\bibinfo {author} {\bibfnamefont {M.}~\bibnamefont
			{Lu}}, \bibinfo {author} {\bibfnamefont {N.~Q.}\ \bibnamefont {Burdick}},
		\bibinfo {author} {\bibfnamefont {S.~H.}\ \bibnamefont {Youn}}, \ and\
		\bibinfo {author} {\bibfnamefont {B.~L.}\ \bibnamefont {Lev}},\ }\href
	{\doibase 10.1103/PhysRevLett.107.190401} {\bibfield  {journal} {\bibinfo
			{journal} {Phys. Rev. Lett.}\ }\textbf {\bibinfo {volume} {107}},\ \bibinfo
		{pages} {190401} (\bibinfo {year} {2011})}\BibitemShut {NoStop}%
	\bibitem [{\citenamefont {Lu}\ \emph {et~al.}(2012)\citenamefont {Lu},
		\citenamefont {Burdick},\ and\ \citenamefont {Lev}}]{lu:quantum_Dy_2012}%
	\BibitemOpen
	\bibfield  {author} {\bibinfo {author} {\bibfnamefont {M.}~\bibnamefont
			{Lu}}, \bibinfo {author} {\bibfnamefont {N.~Q.}\ \bibnamefont {Burdick}}, \
		and\ \bibinfo {author} {\bibfnamefont {B.~L.}\ \bibnamefont {Lev}},\ }\href
	{\doibase 10.1103/PhysRevLett.108.215301} {\bibfield  {journal} {\bibinfo
			{journal} {Phys. Rev. Lett.}\ }\textbf {\bibinfo {volume} {108}},\ \bibinfo
		{pages} {215301} (\bibinfo {year} {2012})}\BibitemShut {NoStop}%
	\bibitem [{\citenamefont {Aikawa}\ \emph {et~al.}(2012)\citenamefont {Aikawa},
		\citenamefont {Frisch}, \citenamefont {Mark}, \citenamefont {Baier},
		\citenamefont {Rietzler}, \citenamefont {Grimm},\ and\ \citenamefont
		{Ferlaino}}]{aikawa:bose-einstein_Er_2012}%
	\BibitemOpen
	\bibfield  {author} {\bibinfo {author} {\bibfnamefont {K.}~\bibnamefont
			{Aikawa}}, \bibinfo {author} {\bibfnamefont {A.}~\bibnamefont {Frisch}},
		\bibinfo {author} {\bibfnamefont {M.}~\bibnamefont {Mark}}, \bibinfo {author}
		{\bibfnamefont {S.}~\bibnamefont {Baier}}, \bibinfo {author} {\bibfnamefont
			{A.}~\bibnamefont {Rietzler}}, \bibinfo {author} {\bibfnamefont
			{R.}~\bibnamefont {Grimm}}, \ and\ \bibinfo {author} {\bibfnamefont
			{F.}~\bibnamefont {Ferlaino}},\ }\href {\doibase
		10.1103/PhysRevLett.108.210401} {\bibfield  {journal} {\bibinfo  {journal}
			{Phys. Rev. Lett.}\ }\textbf {\bibinfo {volume} {108}},\ \bibinfo {pages}
		{210401} (\bibinfo {year} {2012})}\BibitemShut {NoStop}%
	\bibitem [{\citenamefont {Aikawa}\ \emph {et~al.}(2014)\citenamefont {Aikawa},
		\citenamefont {Frisch}, \citenamefont {Mark}, \citenamefont {Baier},
		\citenamefont {Grimm},\ and\ \citenamefont {Ferlaino}}]{Aikawa-Fermi}%
	\BibitemOpen
	\bibfield  {author} {\bibinfo {author} {\bibfnamefont {K.}~\bibnamefont
			{Aikawa}}, \bibinfo {author} {\bibfnamefont {A.}~\bibnamefont {Frisch}},
		\bibinfo {author} {\bibfnamefont {M.}~\bibnamefont {Mark}}, \bibinfo {author}
		{\bibfnamefont {S.}~\bibnamefont {Baier}}, \bibinfo {author} {\bibfnamefont
			{R.}~\bibnamefont {Grimm}}, \ and\ \bibinfo {author} {\bibfnamefont
			{F.}~\bibnamefont {Ferlaino}},\ }\href {\doibase
		10.1103/PhysRevLett.112.010404} {\bibfield  {journal} {\bibinfo  {journal}
			{Phys. Rev. Lett.}\ }\textbf {\bibinfo {volume} {112}},\ \bibinfo {pages}
		{010404} (\bibinfo {year} {2014})}\BibitemShut {NoStop}%
	\bibitem [{\citenamefont {{Brammer}}\ \emph {et~al.}(2012)\citenamefont
		{{Brammer}}, \citenamefont {{Ulitzsch}}, \citenamefont {{Bourouis}},\ and\
		\citenamefont {{Weitz}}}]{brammer}%
	\BibitemOpen
	\bibfield  {author} {\bibinfo {author} {\bibfnamefont {H.}~\bibnamefont
			{{Brammer}}}, \bibinfo {author} {\bibfnamefont {J.}~\bibnamefont
			{{Ulitzsch}}}, \bibinfo {author} {\bibfnamefont {R.}~\bibnamefont
			{{Bourouis}}}, \ and\ \bibinfo {author} {\bibfnamefont {M.}~\bibnamefont
			{{Weitz}}},\ }\href {\doibase 10.1007/s00340-011-4701-2} {\bibfield
		{journal} {\bibinfo  {journal} {Appl. Phys. B}\
		}\textbf {\bibinfo {volume} {106}},\ \bibinfo {pages} {405} (\bibinfo {year}
		{2012})}\BibitemShut {NoStop}%
	\bibitem [{\citenamefont {Frisch}\ \emph {et~al.}(2013)\citenamefont {Frisch},
		\citenamefont {Aikawa}, \citenamefont {Mark}, \citenamefont {Ferlaino},
		\citenamefont {Berseneva},\ and\ \citenamefont {Kotochigova}}]{frisch}%
	\BibitemOpen
	\bibfield  {author} {\bibinfo {author} {\bibfnamefont {A.}~\bibnamefont
			{Frisch}}, \bibinfo {author} {\bibfnamefont {K.}~\bibnamefont {Aikawa}},
		\bibinfo {author} {\bibfnamefont {M.}~\bibnamefont {Mark}}, \bibinfo {author}
		{\bibfnamefont {F.}~\bibnamefont {Ferlaino}}, \bibinfo {author}
		{\bibfnamefont {E.}~\bibnamefont {Berseneva}}, \ and\ \bibinfo {author}
		{\bibfnamefont {S.}~\bibnamefont {Kotochigova}},\ }\href {\doibase
		10.1103/PhysRevA.88.032508} {\bibfield  {journal} {\bibinfo  {journal} {Phys.
				Rev. A}\ }\textbf {\bibinfo {volume} {88}},\ \bibinfo {pages} {032508}
		(\bibinfo {year} {2013})}\BibitemShut {NoStop}%
	\bibitem [{\citenamefont {Shirley}(1982)}]{Shirley:82}%
	\BibitemOpen
	\bibfield  {author} {\bibinfo {author} {\bibfnamefont {J.~H.}\ \bibnamefont
			{Shirley}},\ }\href {\doibase 10.1364/OL.7.000537} {\bibfield  {journal}
		{\bibinfo  {journal} {Opt. Lett.}\ }\textbf {\bibinfo {volume} {7}},\
		\bibinfo {pages} {537} (\bibinfo {year} {1982})}\BibitemShut {NoStop}%
	\bibitem [{\citenamefont {Wybourne}\ and\ \citenamefont
		{Smentek}(2007)}]{Wybourne}%
	\BibitemOpen
	\bibfield  {author} {\bibinfo {author} {\bibfnamefont {B.}~\bibnamefont
			{Wybourne}}\ and\ \bibinfo {author} {\bibfnamefont {L.}~\bibnamefont
			{Smentek}},\ }\href {https://books.google.com/books?id=oPylGiKGQ7UC} {\emph
		{\bibinfo {title} {Optical Spectroscopy of Lanthanides: Magnetic and
				Hyperfine Interactions}}}\ (\bibinfo  {publisher} {CRC Press},\ \bibinfo
	{year} {2007})\BibitemShut {NoStop}%
	\bibitem [{\citenamefont {Baier}\ \emph {et~al.}(2016)\citenamefont {Baier},
		\citenamefont {Mark}, \citenamefont {Petter}, \citenamefont {Aikawa},
		\citenamefont {Chomaz}, \citenamefont {Cai}, \citenamefont {Baranov},
		\citenamefont {Zoller},\ and\ \citenamefont
		{Ferlaino}}]{Ferlaino-Extended-2015}%
	\BibitemOpen
	\bibfield  {author} {\bibinfo {author} {\bibfnamefont {S.}~\bibnamefont
			{Baier}}, \bibinfo {author} {\bibfnamefont {M.~J.}\ \bibnamefont {Mark}},
		\bibinfo {author} {\bibfnamefont {D.}~\bibnamefont {Petter}}, \bibinfo
		{author} {\bibfnamefont {K.}~\bibnamefont {Aikawa}}, \bibinfo {author}
		{\bibfnamefont {L.}~\bibnamefont {Chomaz}}, \bibinfo {author} {\bibfnamefont
			{Z.}~\bibnamefont {Cai}}, \bibinfo {author} {\bibfnamefont {M.}~\bibnamefont
			{Baranov}}, \bibinfo {author} {\bibfnamefont {P.}~\bibnamefont {Zoller}}, \
		and\ \bibinfo {author} {\bibfnamefont {F.}~\bibnamefont {Ferlaino}},\ }\href
	{\doibase 10.1126/science.aac9812} {\bibfield  {journal} {\bibinfo  {journal}
			{Science}\ }\textbf {\bibinfo {volume} {352}},\ \bibinfo {pages} {201}
		(\bibinfo {year} {2016})}\BibitemShut {NoStop}%
	\bibitem [{\citenamefont {de~Paz}\ \emph {et~al.}(2013)\citenamefont {de~Paz},
		\citenamefont {Sharma}, \citenamefont {Chotia}, \citenamefont {Mar\'echal},
		\citenamefont {Huckans}, \citenamefont {Pedri}, \citenamefont {Santos},
		\citenamefont {Gorceix}, \citenamefont {Vernac},\ and\ \citenamefont
		{Laburthe-Tolra}}]{dePaz-Chrom-QM-2013}%
	\BibitemOpen
	\bibfield  {author} {\bibinfo {author} {\bibfnamefont {A.}~\bibnamefont
			{de~Paz}}, \bibinfo {author} {\bibfnamefont {A.}~\bibnamefont {Sharma}},
		\bibinfo {author} {\bibfnamefont {A.}~\bibnamefont {Chotia}}, \bibinfo
		{author} {\bibfnamefont {E.}~\bibnamefont {Mar\'echal}}, \bibinfo {author}
		{\bibfnamefont {J.~H.}\ \bibnamefont {Huckans}}, \bibinfo {author}
		{\bibfnamefont {P.}~\bibnamefont {Pedri}}, \bibinfo {author} {\bibfnamefont
			{L.}~\bibnamefont {Santos}}, \bibinfo {author} {\bibfnamefont
			{O.}~\bibnamefont {Gorceix}}, \bibinfo {author} {\bibfnamefont
			{L.}~\bibnamefont {Vernac}}, \ and\ \bibinfo {author} {\bibfnamefont
			{B.}~\bibnamefont {Laburthe-Tolra}},\ }\href {\doibase
		10.1103/PhysRevLett.111.185305} {\bibfield  {journal} {\bibinfo  {journal}
			{Phys. Rev. Lett.}\ }\textbf {\bibinfo {volume} {111}},\ \bibinfo {pages}
		{185305} (\bibinfo {year} {2013})}\BibitemShut {NoStop}%
	\bibitem [{\citenamefont {Pearman}\ \emph {et~al.}(2002)\citenamefont
		{Pearman}, \citenamefont {Adams}, \citenamefont {Cox}, \citenamefont
		{Griffin}, \citenamefont {Smith},\ and\ \citenamefont {Hughes}}]{pearman}%
	\BibitemOpen
	\bibfield  {author} {\bibinfo {author} {\bibfnamefont {C.~P.}\ \bibnamefont
			{Pearman}}, \bibinfo {author} {\bibfnamefont {C.~S.}\ \bibnamefont {Adams}},
		\bibinfo {author} {\bibfnamefont {S.~G.}\ \bibnamefont {Cox}}, \bibinfo
		{author} {\bibfnamefont {P.~F.}\ \bibnamefont {Griffin}}, \bibinfo {author}
		{\bibfnamefont {D.~A.}\ \bibnamefont {Smith}}, \ and\ \bibinfo {author}
		{\bibfnamefont {I.~G.}\ \bibnamefont {Hughes}},\ }\href
	{http://stacks.iop.org/0953-4075/35/i=24/a=315} {\bibfield  {journal}
		{\bibinfo  {journal} {J. Phys. B: At. Mol. Opt. Phys.}\ }\textbf {\bibinfo
			{volume} {35}},\ \bibinfo {pages} {5141} (\bibinfo {year}
		{2002})}\BibitemShut {NoStop}%
	\bibitem [{\citenamefont {Harris}\ \emph {et~al.}(2006)\citenamefont {Harris},
		\citenamefont {Adams}, \citenamefont {Cornish}, \citenamefont {McLeod},
		\citenamefont {Tarleton},\ and\ \citenamefont {Hughes}}]{harris}%
	\BibitemOpen
	\bibfield  {author} {\bibinfo {author} {\bibfnamefont {M.~L.}\ \bibnamefont
			{Harris}}, \bibinfo {author} {\bibfnamefont {C.~S.}\ \bibnamefont {Adams}},
		\bibinfo {author} {\bibfnamefont {S.~L.}\ \bibnamefont {Cornish}}, \bibinfo
		{author} {\bibfnamefont {I.~C.}\ \bibnamefont {McLeod}}, \bibinfo {author}
		{\bibfnamefont {E.}~\bibnamefont {Tarleton}}, \ and\ \bibinfo {author}
		{\bibfnamefont {I.~G.}\ \bibnamefont {Hughes}},\ }\href {\doibase
		10.1103/PhysRevA.73.062509} {\bibfield  {journal} {\bibinfo  {journal} {Phys.
				Rev. A}\ }\textbf {\bibinfo {volume} {73}},\ \bibinfo {pages} {062509}
		(\bibinfo {year} {2006})}\BibitemShut {NoStop}%
	\bibitem [{\citenamefont {Yoshikawa}\ \emph {et~al.}(2003)\citenamefont
		{Yoshikawa}, \citenamefont {Umeki}, \citenamefont {Mukae}, \citenamefont
		{Torii},\ and\ \citenamefont {Kuga}}]{Yoshikawa:03}%
	\BibitemOpen
	\bibfield  {author} {\bibinfo {author} {\bibfnamefont {Y.}~\bibnamefont
			{Yoshikawa}}, \bibinfo {author} {\bibfnamefont {T.}~\bibnamefont {Umeki}},
		\bibinfo {author} {\bibfnamefont {T.}~\bibnamefont {Mukae}}, \bibinfo
		{author} {\bibfnamefont {Y.}~\bibnamefont {Torii}}, \ and\ \bibinfo {author}
		{\bibfnamefont {T.}~\bibnamefont {Kuga}},\ }\href {\doibase
		10.1364/AO.42.006645} {\bibfield  {journal} {\bibinfo  {journal} {Appl.
				Opt.}\ }\textbf {\bibinfo {volume} {42}},\ \bibinfo {pages} {6645} (\bibinfo
		{year} {2003})}\BibitemShut {NoStop}%
	\bibitem [{\citenamefont {Zhu}\ \emph {et~al.}(2014)\citenamefont {Zhu},
		\citenamefont {Chen}, \citenamefont {Li},\ and\ \citenamefont
		{Wang}}]{Zhu:14}%
	\BibitemOpen
	\bibfield  {author} {\bibinfo {author} {\bibfnamefont {S.}~\bibnamefont
			{Zhu}}, \bibinfo {author} {\bibfnamefont {T.}~\bibnamefont {Chen}}, \bibinfo
		{author} {\bibfnamefont {X.}~\bibnamefont {Li}}, \ and\ \bibinfo {author}
		{\bibfnamefont {Y.}~\bibnamefont {Wang}},\ }\href {\doibase
		10.1364/JOSAB.31.002302} {\bibfield  {journal} {\bibinfo  {journal} {J. Opt.
				Soc. Am. B}\ }\textbf {\bibinfo {volume} {31}},\ \bibinfo {pages} {2302}
		(\bibinfo {year} {2014})}\BibitemShut {NoStop}%
	\bibitem [{\citenamefont {Do}\ \emph {et~al.}(2008)\citenamefont {Do},
		\citenamefont {Moon},\ and\ \citenamefont {Noh}}]{Do}%
	\BibitemOpen
	\bibfield  {author} {\bibinfo {author} {\bibfnamefont {H.~D.}\ \bibnamefont
			{Do}}, \bibinfo {author} {\bibfnamefont {G.}~\bibnamefont {Moon}}, \ and\
		\bibinfo {author} {\bibfnamefont {H.}~\bibnamefont {Noh}},\ }\href {\doibase
		10.1103/PhysRevA.77.032513} {\bibfield  {journal} {\bibinfo  {journal} {Phys.
				Rev. A}\ }\textbf {\bibinfo {volume} {77}},\ \bibinfo {pages} {032513}
		(\bibinfo {year} {2008})}\BibitemShut {NoStop}%
	\bibitem [{\citenamefont {Bennett}(1962)}]{HoleBurning}%
	\BibitemOpen
	\bibfield  {author} {\bibinfo {author} {\bibfnamefont {W.~R.}\ \bibnamefont
			{Bennett}},\ }\href {\doibase 10.1103/PhysRev.126.580} {\bibfield  {journal}
		{\bibinfo  {journal} {Phys. Rev.}\ }\textbf {\bibinfo {volume} {126}},\
		\bibinfo {pages} {580} (\bibinfo {year} {1962})}\BibitemShut {NoStop}%
	\bibitem [{\citenamefont {Smith}\ and\ \citenamefont
		{H\"ansch}(1971)}]{SatSpec}%
	\BibitemOpen
	\bibfield  {author} {\bibinfo {author} {\bibfnamefont {P.~W.}\ \bibnamefont
			{Smith}}\ and\ \bibinfo {author} {\bibfnamefont {R.}~\bibnamefont
			{H\"ansch}},\ }\href {\doibase 10.1103/PhysRevLett.26.740} {\bibfield
		{journal} {\bibinfo  {journal} {Phys. Rev. Lett.}\ }\textbf {\bibinfo
			{volume} {26}},\ \bibinfo {pages} {740} (\bibinfo {year} {1971})}\BibitemShut
	{NoStop}%
	\bibitem [{\citenamefont {Wieman}\ and\ \citenamefont
		{H\"ansch}(1976)}]{WiemanPolSpec}%
	\BibitemOpen
	\bibfield  {author} {\bibinfo {author} {\bibfnamefont {C.}~\bibnamefont
			{Wieman}}\ and\ \bibinfo {author} {\bibfnamefont {T.~W.}\ \bibnamefont
			{H\"ansch}},\ }\href {\doibase 10.1103/PhysRevLett.36.1170} {\bibfield
		{journal} {\bibinfo  {journal} {Phys. Rev. Lett.}\ }\textbf {\bibinfo
			{volume} {36}},\ \bibinfo {pages} {1170} (\bibinfo {year}
		{1976})}\BibitemShut {NoStop}%
	\bibitem [{\citenamefont {Hartog}\ \emph {et~al.}(2010)\citenamefont {Hartog},
		\citenamefont {Chisholm},\ and\ \citenamefont {Lawler}}]{hartog}%
	\BibitemOpen
	\bibfield  {author} {\bibinfo {author} {\bibfnamefont {E.~A.~D.}\
			\bibnamefont {Hartog}}, \bibinfo {author} {\bibfnamefont {J.~P.}\
			\bibnamefont {Chisholm}}, \ and\ \bibinfo {author} {\bibfnamefont {J.~E.}\
			\bibnamefont {Lawler}},\ }\href
	{http://stacks.iop.org/0953-4075/43/i=15/a=155004} {\bibfield  {journal}
		{\bibinfo  {journal} {J. Phys. B: At. Mol. Opt. Phys.}\ }\textbf {\bibinfo
			{volume} {43}},\ \bibinfo {pages} {155004} (\bibinfo {year}
		{2010})}\BibitemShut {NoStop}%
	\bibitem [{\citenamefont {Lawler}\ \emph {et~al.}(2010)\citenamefont {Lawler},
		\citenamefont {Wyart},\ and\ \citenamefont {Hartog}}]{lawler}%
	\BibitemOpen
	\bibfield  {author} {\bibinfo {author} {\bibfnamefont {J.~E.}\ \bibnamefont
			{Lawler}}, \bibinfo {author} {\bibfnamefont {J.-F.}\ \bibnamefont {Wyart}}, \
		and\ \bibinfo {author} {\bibfnamefont {E.~A.~D.}\ \bibnamefont {Hartog}},\
	}\href {http://stacks.iop.org/0953-4075/43/i=23/a=235001} {\bibfield
	{journal} {\bibinfo  {journal} {J. Phys. B: At. Mol. Opt. Phys.}\ }\textbf
	{\bibinfo {volume} {43}},\ \bibinfo {pages} {235001} (\bibinfo {year}
	{2010})}\BibitemShut {NoStop}%
\bibitem [{\citenamefont {Demtr{\"o}der}(2008)}]{demtroederv2}%
\BibitemOpen
\bibfield  {author} {\bibinfo {author} {\bibfnamefont {W.}~\bibnamefont
		{Demtr{\"o}der}},\ }\href {https://books.google.com/books?id=QDvDNAEACAAJ}
{\emph {\bibinfo {title} {Laser Spectroscopy: Vol. 2: Experimental
			Techniques}}}\ (\bibinfo  {publisher} {Springer Berlin Heidelberg},\ \bibinfo
{year} {2008})\BibitemShut {NoStop}%
\bibitem [{\citenamefont {Metcalf}\ and\ \citenamefont {van~der
		Straten}(2001)}]{metcalf}%
\BibitemOpen
\bibfield  {author} {\bibinfo {author} {\bibfnamefont {H.}~\bibnamefont
		{Metcalf}}\ and\ \bibinfo {author} {\bibfnamefont {P.}~\bibnamefont {van~der
			Straten}},\ }\href {https://books.google.com/books?id=i-40VaXqrj0C} {\emph
	{\bibinfo {title} {Laser Cooling and Trapping}}},\ Graduate Texts in
Contemporary Physics\ (\bibinfo  {publisher} {Springer New York},\ \bibinfo
{year} {2001})\BibitemShut {NoStop}%
\end{thebibliography}

%

\end{document}